\documentclass[useAMS,usenatbib]{mn2e}%,doublespacing
\usepackage{graphicx,fleqn,fix2col,rotating}
\DeclareGraphicsExtensions{.eps,.ps,.eps.gz,.ps.gz,.eps.Z}
\title[Physical properties of IP Pegasi]{Physical properties of IP Pegasi: an eclipsing dwarf nova with an unusually cool white dwarf}

\author[C.M.~Copperwheat et al.]{C.M.~Copperwheat$^{1}$, T.R.~Marsh$^{1}$, V.S.~Dhillon$^{2}$, S.P.~Littlefair$^{2}$, \newauthor  R.~Hickman$^{1}$, B.T.~G{\"{a}}nsicke$^{1}$ and J.~Southworth$^{1}$\\
$^{1}$ Department of Physics, University of Warwick, Coventry, CV4 7AL, UK\\
$^{2}$ Department of Physics and Astronomy, University of Sheffield, S3 7RH, UK
}

\date{Received: }

\begin{document}

\newcommand{\dg} {^{\circ}}
\outer\def\gtae {$\buildrel {\lower3pt\hbox{$>$}} \over
{\lower2pt\hbox{$\sim$}} $}
\outer\def\ltae {$\buildrel {\lower3pt\hbox{$<$}} \over
{\lower2pt\hbox{$\sim$}} $}
\newcommand{\ergscm} {erg s$^{-1}$ cm$^{-2}$}
\newcommand{\ergss} {erg s$^{-1}$}
\newcommand{\ergsd} {erg s$^{-1}$ $d^{2}_{100}$}
\newcommand{\pcmsq} {cm$^{-2}$}
\newcommand{\ros} {{\it ROSAT}}
\newcommand{\xmm} {\mbox{{\it XMM-Newton}}}
\newcommand{\exo} {{\it EXOSAT}}
\newcommand{\sax} {{\it BeppoSAX}}
\newcommand{\chandra} {{\it Chandra}}
\newcommand{\hst} {{\it HST}}
\newcommand{\subaru} {{\it Subaru}}
\def\rchi{{${\chi}_{\nu}^{2}$}}
\newcommand{\Msun} {$M_{\odot}$}
\newcommand{\Mwd} {$M_{wd}$}
\newcommand{\Mbh} {$M_{\bullet}$}
\newcommand{\Lsun} {$L_{\odot}$}
\newcommand{\Rsun} {$R_{\odot}$}
\newcommand{\Zsun} {$Z_{\odot}$}
\def\Mdot{\hbox{$\dot M$}}
\def\mdot{\hbox{$\dot m$}}
\def\mincir{\raise -2.truept\hbox{\rlap{\hbox{$\sim$}}\raise5.truept
\hbox{$<$}\ }}
\def\magcir{\raise -4.truept\hbox{\rlap{\hbox{$\sim$}}\raise5.truept
\hbox{$>$}\ }}
\newcommand{\mnras} {MNRAS}
\newcommand{\aap} {A\&A}
\newcommand{\apj} {ApJ}
\newcommand{\apjl} {ApJL}
\newcommand{\apjs} {ApJS}
\newcommand{\aj} {AJ}
\newcommand{\pasp} {PASP}
\newcommand{\apss} {Ap\&SS}
\maketitle

\begin{abstract} 
We present high speed photometric observations of the eclipsing dwarf nova IP Peg taken with the triple-beam camera ULTRACAM mounted on the William Herschel Telescope. The primary eclipse in this system was observed twice in 2004, and then a further sixteen times over a three week period in 2005. Our observations were simultaneous in the Sloan $u'$, $g'$ and $r'$ bands. By phase-folding and averaging our data we make the first significant detection of the white dwarf ingress in this system and find the phase width $\phi$ of the white dwarf eclipse to be $0.0935 \pm 0.0003$, significantly higher than the previous best value of $0.0863 < \phi < 0.0918$. The mass ratio  is found to be $q = M_2 / M_1 = 0.48 \pm 0.01$, consistent with previous measurements, but we find the inclination to be $83.8 \pm 0.5$ deg, significantly higher than previously reported. We find the radius of the white dwarf to be $0.0063 \pm 0.0003$\Rsun, implying a white dwarf mass of $1.16 \pm 0.02$\Msun. The donor mass is $0.55 \pm 0.02$\Msun. The white dwarf temperature is more difficult to determine, since the white dwarf is seen to vary significantly in flux, even between consecutive eclipses. This is seen particularly in the $u'$-band, and is probably the result of absorption by disc material. Our best estimate of the temperature is $10,000$ -- $15,000$K, which is much lower than would be expected for a CV with this period, and implies a mean accretion rate of $< 5 \times 10^{-11} M_{\odot}$ yr$^{-1}$, more than $40$ times lower than the expected rate.
\end{abstract}


%%%%%%%%%%%%%%%%%%%%%%%%%% Begin Section 1 %%%%%%%%%%%%%%%%%%%%%%%%%%%%%
\section{INTRODUCTION}  

Cataclysmic variable stars (CVs: \citealt{Warner95}) provide examples of white dwarfs accreting from low mass companions at rates of $\sim$$10^{-11}$ -- $10^{-9}$\Msun/yr. The mass transfer in CVs and similar semi-detached binary systems is driven by angular momentum loss. For long orbital period ($\sim$$3$ h or greater) systems the angular momentum loss is thought to be driven by magnetically-coupled stellar winds. For the shorter period ($< 3$ h) systems losses due to  gravitational radiation are thought to dominate the mass transfer process. This is observationally supported to some extent: the shorter period population is dominated by lower \Mdot \ dwarf novae, whereas the peak of the population of high \Mdot \ `steady state' CVs is in the $3$ to $4$h period range \citep{Shafter92}. However, there is no one-to-one relationship. The  distribution of steady-state, bright CVs  tails off sharply at longer periods \citep{Ritter03}, and a number of dwarf novae  are observed to have very long periods. It is generally thought that all non-magnetic CVs are essentially the same, with only a change in \Mdot \ required to move between the different subclasses. While it may be the case that the high- and low-\Mdot \ systems at a given orbital period are different populations with a different evolutionary history, a more likely explanation is that the instantaneous \Mdot \ we measure is a poor indicator of the mean \Mdot, so that even high mean \Mdot \ systems spend some fraction of their time as slowly accreting dwarf novae. 

One independent means of inferring a longer-term ($10^3$ -- $10^5$yr) value of \Mdot \  is through measurement of the white dwarf temperature. The temperature is a good tracer of the long-term \Mdot \ since it is determined not by the accretion heating, but by the compression of the underlying white dwarf by the accreted matter \citep{Townsley03, Townsley09}. The existing measurements agree well with the high-\Mdot/long period, low-\Mdot/short period consensus (figure 5 of \citealt{Townsley09}). However, these data are likely to be biased. These white dwarf temperatures are spectroscopically determined, and the white dwarf must dominate at UV wavelengths in order for the temperature to be determined. Long period CVs are generally brighter than shorter period systems due to their higher accretion rates and more luminous donor stars, and so the white dwarfs in long period systems must be hotter in order to be measurable. An alternative strategy which reduces this bias is to make fast, multi wavelength photometric observations of CVs where the inclination is high enough for the white dwarf to be eclipsed. With sufficient time resolution the ingress and egress of the white dwarf can be identified independently of the other components, and hence the white dwarf colours can be determined.

 An ideal system for a study of this nature is IP Pegasi (IP Peg hereafter). This dwarf nova was found to be eclipsing by \citet{Goranskii85a} and has a period of $3.8$h, in the `long period' tail of the dwarf nova distribution. Following \citet{Townsley03} this would imply a high mean \Mdot \ and a correspondingly high white dwarf temperature. However, the temperature of the white dwarf has yet to be determined with precision. Previous studies of IP Peg have shown that the white dwarf ingress feature is obscured by the ingress of the bright spot on the accretion disc, and it has been claimed that the white dwarf egress feature varies considerably in size and length \citep{Wood86, Wood89}. The difficulty in detecting the white dwarf throws the system parameters of IP Peg into doubt. In an attempt to resolve this and to measure the temperature of the white dwarf we took a total of eighteen separate observations of the IP Peg eclipse with the high speed CCD camera ULTRACAM mounted on the William Herschel Telescope (WHT). In this paper we present these data, and our subsequent parameter determinations.

\section{OBSERVATIONS}
\label{sec:obs}

\begin{table*} 
\caption{Log of the observations. Each observation is a single eclipse of IP Peg. Unless otherwise stated, the Sloan $u'$, $g'$ and $r'$ filters were used. We give the phase range centred on the white dwarf egress.}
\label{tab:obs} 
\begin{tabular}{llllllll} 
Eclipse     &           &Date           &\multicolumn{2}{c}{UT} &Exposure		&Phase			&\\
number      &Cycle      &start  	    &start	    	&end    &time (s)	    &range		&Comments\\
\hline
1           &$-2200$    &29 Aug 2004	&$01:10$	&$06:05$	&$1.5$		    &$-0.60$ -- $0.70$	&Clear, seeing $0.6$ -- $0.8$''. \\
            &           &               &           &           &               &                   &$i'$-band filter used in place of $r'$.\\
2           &$-2194$    &30 Aug	        &$02:04$	&$02:24$	&$1.5$		    &$-0.06$ -- $0.02$	&Clear, seeing $0.6$ -- $0.8$''. \\
            &           &               &           &           &               &                   &Partial lightcurve: White dwarf egress only.\\
3           &$-1$       &12 Aug 2005	&$00:41$	&$01:17$	&$1.5$		    &$-0.09$ -- $0.07$	&Clear, seeing $0.6$ -- $1.0$''.\\
            &           &               &           &           &               &                   &M-dwarf flare just before white dwarf egress.\\
4           &$0$        &12 Aug	        &$04:17$	&$05:47$	&$2.5$		    &$-0.14$ -- $0.24$	&Mostly clear, seeing $0.6$''.\\
5           &$6$        &13 Aug	        &$03:06$	&$04:00$	&$1.5$		    &$-0.10$ -- $0.10$	&Clear, seeing $0.6$''. More readout noise than\\ 
            &           &		        &		    &		    &		        &			        &usual in the $r'$-band due to a CCD problem.\\
6           &$18$       &15 Aug	        &$00:33$	&$01:34$	&$1.5$		    &$-0.16$ -- $0.10$	&Clear, seeing $0.6$''.\\
7           &$19$       &15 Aug	        &$04:20$	&$05:31$	&$1.5$		    &$-0.03$ -- $0.14$	&Clear, seeing $0.6$''.\\
            &           &		        &		    &		    &		        &			        &Partial lightcurve: white dwarf egress only.\\
8           &$24$       &15 Aug	        &$23:17$	&$00:17$	&$1.5$		    &$-0.17$ -- $0.08$	&Clear, seeing $0.6$''.\\
9           &$87$       &25 Aug	        &$22:28$	&$23:27$	&$1.5$		    &$-0.08$ -- $0.08$	&Partial lightcurve due to cloud. \\
            &           &		        &		    &		    &		        &			        &Seeing $0.6$ -- $1.0$''. Dust.\\
10          &$88$       &26 Aug	        &$01:47$	&$03:18$	&$1.5$		    &$-0.18$ -- $0.09$	&Partial lightcurve due to cloud. \\
            &           &		        &		    &		    &		        &			        &Seeing $0.6$ -- $1.0$''. Dust.\\
11          &$101$      &28 Aug	        &$03:30$	&$04:35$	&$0.9$		    &$-0.21$ -- $0.07$	&Seeing $0.5$''. Dust.\\
12          &$106$      &28 Aug	        &$22:43$	&$23:38$	&$1.2$		    &$-0.15$ -- $0.08$	&Seeing $1$ --$2$''. Some dust.\\
13          &$107$      &29 Aug	        &$02:24$	&$03:21$	&$1.2$		    &$-0.17$ -- $0.07$	&Seeing $1$ --$2$''. Some dust.\\
14          &$112$      &29 Aug	        &$21:47$	&$22:18$	&$2.5$		    &$-0.07$ -- $0.06$	&Seeing $1$ --$3$''. Dust.\\
            &  &		&		        &		    &		    &			                        &Partial lightcurve -- no ingress.\\
15          &$113$      &30 Aug	        &$01:18$	&$02:10$	&$1.5$		    &$-0.14$ -- $0.08$	&Seeing $1$ -- $2$''. Dust.\\
16          &$119$      &31 Aug	        &$00:02$	&$00:57$	&$0.9$		    &$-0.16$ -- $0.08$	&Seeing $\sim$$0.8$''. Dust.\\
17          &$120$      &31 Aug	        &$03:50$	&$04:43$	&$0.9$		    &$-0.16$ -- $0.07$	&Seeing $0.5$ -- $0.6$''. Much dust.\\
18          &$133$      &02 Sept	    &$04:25$	&$06:11$	&$1.2$ -- $2.0$ &$-0.36$ -- $0.08$	&Seeing $1$ -- $2$''. Dust.\\
\hline
\end{tabular}
\end{table*}

The high speed CCD camera ULTRACAM \citep{Dhillon07} was mounted on the $4.2$m William Herschel Telescope (WHT) in August $2005$. Observations were made with this instrument between $9$ -- $15$ August, and $25$ August -- $1$ September. Over these periods, $16$ separate observations of the IP Peg eclipse were made. In addition, two observations of the eclipse were made in August 2004. ULTRACAM is a triple beam camera and all observations were made using the SDSS $u'$, $g'$ and $r'$ filters, except for the first 2004 observation, in which the $i'$ filter was used in place of $r'$, for scheduling reasons. Most observations were taken with an exposure time of $1.5$s, which gave a sufficient count rate in all three bands and adequately sampled the white dwarf egress, which lasts $\sim$$30$s. The dead time between exposures for ULTRACAM is $\sim$$25$ms. The data were unbinned and the CCD was windowed in order to achieve this exposure time. This exposure time was varied by up to $1$s in order to account for changing conditions. A complete log of the observations is given in Table \ref{tab:obs}.

All of these data were reduced with aperture photometry using the ULTRACAM pipeline software, with debiassing, flatfielding and sky background subtraction performed in the standard way. The source flux was determined using a variable aperture (whereby the radius of the aperture is scaled according to the FWHM by a factor of $1.7$). Variation in observing conditions were accounted for by dividing the source lightcurve by the lightcurve of a nearby comparison star. We use the star at $\alpha = 23h 23m 06s$, $\delta = +18^{\circ} 24' 40''$ as the comparison. We checked the stability of this star against other stars in the field, and calculated apparent magnitudes of $14.45$, $13.20$ and $12.62$ in $u'$, $g'$ and $r'$ respectively.

Our short eclipse observations were interleaved with longer observations for other programmes. We used the field stars in these long datasets to determine atmospheric absorption coefficients in the $u'$, $g'$ and $r'$ bands, and subsequently determined the absolute flux of our targets using observations of standard stars (from \citealt{Smith02}) taken in evening twilight. We use this calibration for our determinations of the apparent magnitudes of the two sources, although we present all lightcurves in flux units determined using the conversion given in \citet{Smith02}. Using our absorption coefficients we extrapolate all fluxes to an airmass of $0$. For all data we convert the MJD times to the barycentric dynamical timescale, applying light travel times to the solar system barycentre.

\section{LIGHTCURVES}
\label{sec:lightcurves}

\begin{figure*}
\centering
\includegraphics[angle=270,width=1.0\textwidth]{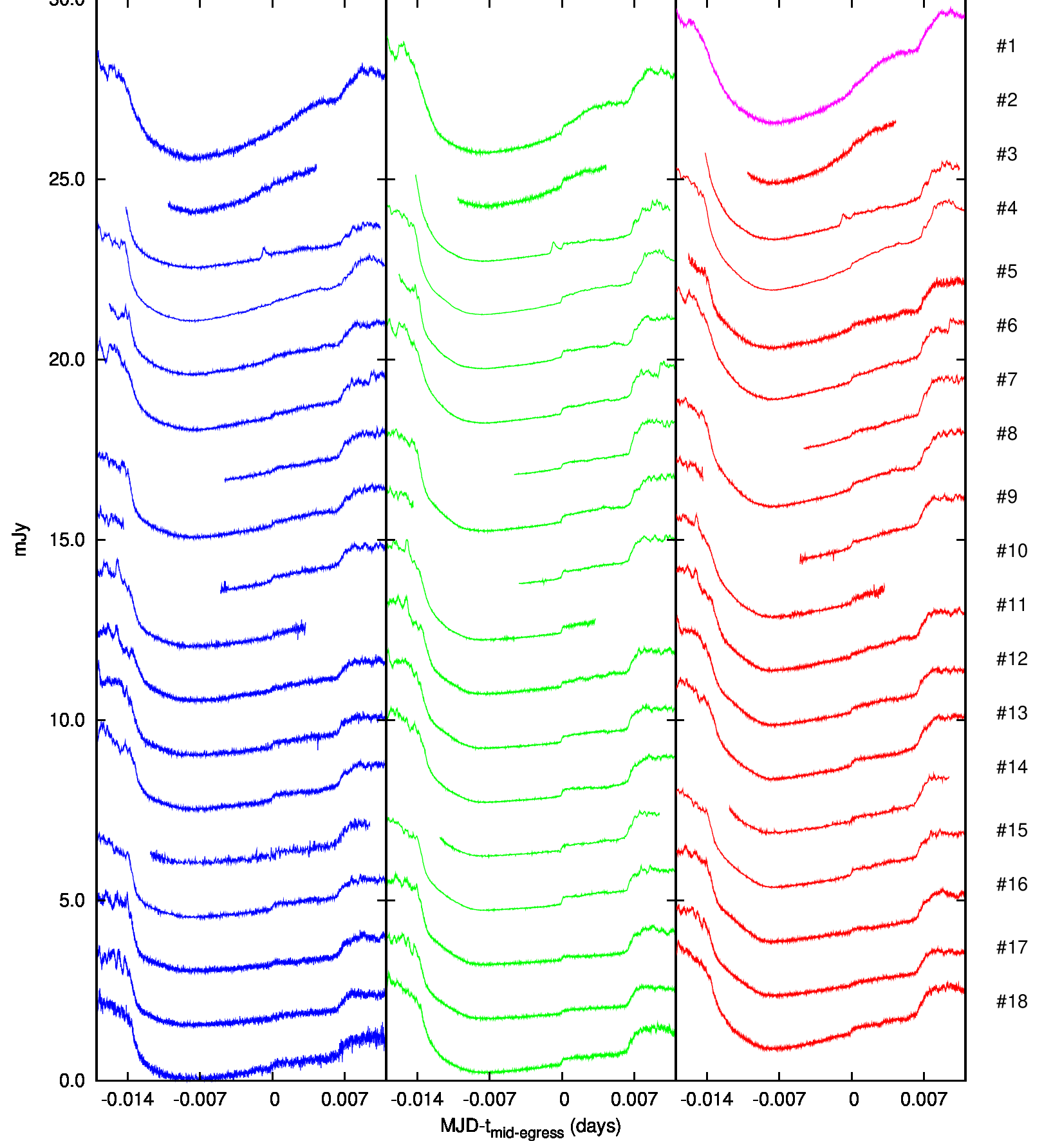}
\hfill
\caption{Lightcurves of IP Peg eclipses, observed in August 2004 and August 2005. We label the eclipses according to the numbers given in the observation log (Table \ref{tab:obs}). All data were collected simultaneously in the $u'$- $g'$- and $r'$-bands (blue, green and red lines, respectively) except for eclipse $1$, in which an $i'$ filter was used in place of $r'$. Each lightcurve in this plot is offset from the previous by $1.5$mJy.} \label{fig:lightcurves_1} \end{figure*}

In Figure \ref{fig:lightcurves_1} the eighteen eclipses that were observed in $2004$ and $2005$ are plotted. In some cases we only have a partial lightcurve, due to either poor weather or scheduling constraints. We comment on weather conditions in Table \ref{tab:obs}. In particular, note that the observations made on $25$ August and later (eclipses $9$ to $18$) suffered from high levels of Saharan dust in the atmosphere. This dust has a grey absorption curve but the loss of flux is particularly apparent in the $u'$-band lightcurves.

These lightcurves show a single ingress feature, followed by separate egresses of the white dwarf, and later, the bright spot. While the egresses of these two components are distinct, it is impossible in these plots to disentangle the two ingresses. There is considerable accretion-driven flickering out of eclipse, most of which is due to the bright spot, although in some cases there is a degree of variation immediately after the white dwarf egress. There is noticeable variation in the strength of the white dwarf egress feature. This will be addressed in detail in Section \ref{sec:entireeclipse}.
The $2004$ observations are at a different point in the outburst cycle to the $2005$ observations, and so the shape of eclipses $1$ and $2$ are different to the rest, due to the accretion disc flux. In addition, in eclipse $1$ there is a `hump' between the white dwarf and bright spot egresses, which may be due to flickering. Finally, it can be seen in eclipse number $3$ that there is a brief brightening in the lightcurve just prior to the white dwarf egress. This brightening is most likely a flare event from the donor star, and makes it impossible to model the egress for that eclipse accurately.

\section{ECLIPSE MODELLING}
\label{sec:model}

In this section we report system parameters determined by fitting the data with the model described in Appendix \ref{sec:appendix}. This was a three stage process. We first obtained an initial fit to each individual eclipse using the simplex and Levenberg-Marquardt methods \citep{Press02}. These initial fits were used to obtain eclipse timings, and the resultant ephemeris was used to phase-fold all of the complete eclipse lightcurves. This combined lightcurve was used to determine the mass ratio $q$ and inclination $i$ of the system. We use a Markov Chain Monte Carlo (MCMC) algorithm for minimisation and determination of uncertainties. These parameters were used as a basis for MCMC fits of the individual lightcurves, so as to determine parameters which would be expected to vary over the course of our observations, such as the accretion disc radius. Since the white dwarf egress feature appears to vary in height, and since there have been reports of variable egress duration \citep{Wood86,Wood89}, the white dwarf flux and radius were determined individually for each eclipse, as well as more refined eclipse timings which we used to determine a new ephemeris for this system.

\subsection{Fitting the phase-folded data}
\label{sec:phasefold}

\begin{figure*}
\centering
\includegraphics[angle=270,width=1.0\textwidth]{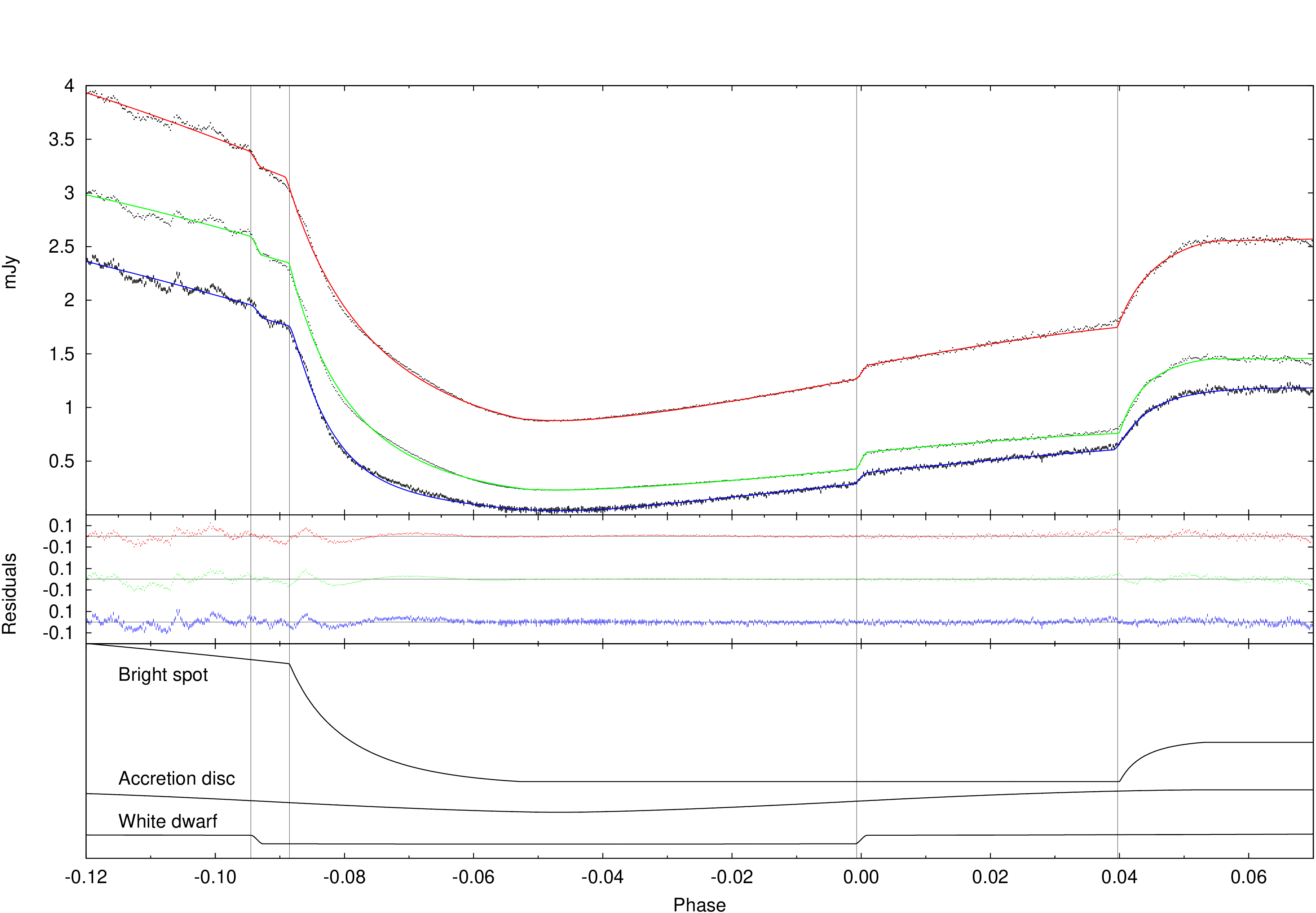}
\hfill
\caption{Top panel: phase-folded and binned lightcurves (top, $r'$; middle, $g'$; bottom, $u'$), comprised of the eleven complete eclipse observations made in 2004 and 2005. We plot the average flux in mJy against the binary phase, where a phase of $0$ corresponds to the mid-egress of the white dwarf. We plot the datapoints with uncertainties in black, and the best model fits to these data in red, green and blue (for $r'$, $g'$ and $u'$, respectively). The four vertical lines indicate the start of (from left to right) the white dwarf ingress, the bright spot ingress, the white dwarf egress and the bright spot egress. Middle panel: residuals in the three bands (top, $r'$; middle, $g'$; bottom, $u'$). Bottom panel: the three components of the $g'$-band model plotted separately, showing the relative strengths of the bright spot, accretion disc and white dwarf. These three lines have been offset for clarity.} \label{fig:phasefold} \end{figure*}

We selected the eleven lightcurves in 2005 where the entire eclipse is covered (numbers $4$, $5$, $6$, $8$, $11$, $12$, $13$, $15$, $16$, $17$ and $18$), and phase-folded them using the ephemeris given in Section \ref{sec:timing}. The combined and binned lightcurves are plotted in Figure \ref{fig:phasefold}, using all eleven complete eclipses. Each bin is $\sim$$2.7$s in length. With the flickering reduced, a white dwarf ingress feature is clearly apparent, distinct from and preceding the ingress of the bright spot. We measured the phase width of the white dwarf eclipse to be $0.0935 \pm 0.0003$, which is greater than the value of $0.0863$ suggested as being most likely by \citet{Wood86} and is also greater than the upper limit of $0.0918$ given in the same paper. This implies either a higher orbital inclination or a higher mass ratio or some combination of the two.

It might be thought that this feature could be due to flickering. However, in all three bands it has a height and duration that is consistent with it being the white dwarf ingress. Additionally, by changing the number of lightcurves used to create the phase-folded plot, we find that the existence and size of this feature in the combined dataset is not due to any one individual lightcurve. This is not true of other significant variations in the lightcurve, such as the two peaks immediately preceding the ingress at a phase range of $-0.11 < \phi < -0.10$, which are definitely due to flickering. We are therefore confident that we have detected the ingress feature.

A model consisting of a white dwarf primary, a Roche lobe filling secondary and accretion disc and bright spot components was fitted to these data using the MCMC method. A complete description of this model is given in Appendix \ref{sec:appendix}. In Table \ref{tab:mcmcresults} we list the parameter determinations we make with this method, although some of these are refit on an eclipse-by-eclipse basis in Section \ref{sec:entireeclipse}. The uncertainties listed in this table are the formal errors, scaled so as to give a reduced $\chi^2$ of $1$. These uncertainties are underestimates, as shown by the inconsistencies between the three photometric bands. In Section \ref{sec:massratio} we will discuss this issue further, and calculate more realistic uncertainties which  account for the systematics. The best model fits are plotted in Figure \ref{fig:phasefold}. The fits are generally good, although there is some variation in the residuals out-of-eclipse due to flickering which has not been completely averaged out in the combined lightcurve. Additionally, our fit to the bright spot ingress and egress is not perfect; we discuss this in Appendix \ref{sec:app_bs}.

One additional parameter which our results are sensitive to is the limb darkening coefficient for the white dwarf. An initial fit to the data was made with the Levenberg-Marquardt method in order to obtain approximate white dwarf parameters, and these determinations were used with the white dwarf model atmospheres of \citet*{Gaensicke95} to determine the coefficients in the three ULTRACAM bands. We found linear coefficients of $0.352$, $0.253$, $0.216$ in $u'$, $g'$ and $r'$ to be appropriate for the white dwarf in IP Peg. These values were used in all our MCMC fits.

\subsection{Fitting the individual eclipses}
\label{sec:entireeclipse}

For some parameters the results from the phase-folded lightcurves are insufficient. Some variations in the white dwarf egress feature are clearly apparent in the individual eclipses, so it was necessary to fit these separately in order to obtain a proper determination of the white dwarf radius and fluxes. Obviously this was also necessary in order to determine the eclipse timings, and some variation in the accretion disc radius over the course of our observations would also be expected. We used our model fit to the phase-folded lightcurve as a basis for fitting the individual eclipses separately. It is clear from Figure \ref{fig:lightcurves_1} that the amplitude of the flickering is much larger than the white dwarf eclipse features in these individual lightcurves, so we fixed most of the parameters, including the mass ratio and inclination angle, to the values determined from the phase-folded lightcurve. We allowed the white dwarf radius and the time of the white dwarf mid-eclipse to vary. We also wanted to track variations in the disc radius, but in these individual lightcurves the outer disc radius $R_{disc}$ is poorly constrained. We therefore set  $R_{disc} = R_{spot}$ and fitted $R_{spot}$, where $R_{spot}$ is the distance between the white dwarf and the maximum of the surface brightness of the bright spot. This parameter is well constrained by the data, since the bright spot is so strong in flux. Note that in the previous section (the phase-folded lightcurves) the data are of sufficient quality that we could fit $R_{disc}$ and $R_{spot}$ separately, and we found $R_{spot}$ to track the accretion disc radius closely. To allow for a degree of flexibility in modelling a variable bright spot, we also allow $l$ (the scale length of the bright spot) and $f_c$ (the fraction of bright spot flux that is taken to be constant) to vary. The remaining bright spot parameters were fixed to the values determined from the phase-folded lightcurve. Finally the surface brightness of the accretion disc is modelled as a power law, and the exponent of this power law is also allowed to vary. A complete description of all parameters is given in Appendix \ref{sec:appendix}. Each individual lightcurve was fitted in the same way as the phase-folded lightcurve (as described in Section \ref{sec:phasefold}). 

\begin{figure*}
\centering
\includegraphics[angle=270,width=1.0\textwidth]{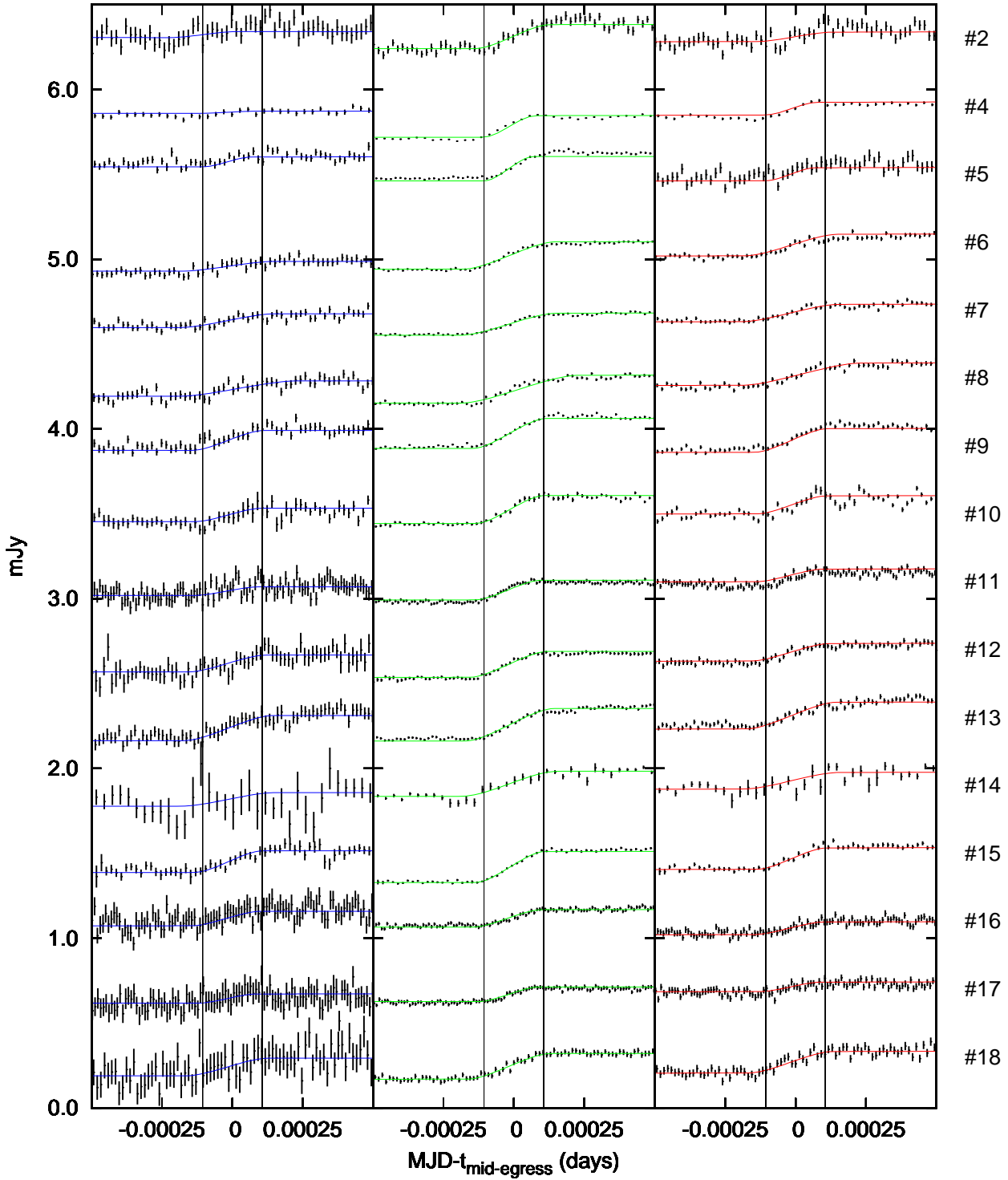}
\hfill
\caption{The region of the lightcurves around the white dwarf egress for sixteen of the eighteen eclipses. We label the eclipses according to the numbers given in the observation log (Table \ref{tab:obs}). We plot the $u'$-band data in the first column, the $g'$-band data in the second and the $r'$-band data in the third. For the x-axis timings we used the ephemeris given in Section \ref{sec:timing}. The dashed vertical lines identify the mean phase width of the white dwarf egress.} \label{fig:egress} \end{figure*}

The results for the $g'$-band fits (for which the signal-to-noise is highest) are listed in Table \ref{tab:wdparams}. We report the white dwarf radii and timings for seventeen of the eighteen eclipses, omitting the eclipse with the flare near the egress feature. There is a considerable degree of scatter in our model determinations of the white dwarf radius. However, given the uncertainties, the variations are not significant. The timings of the white dwarf mid-eclipse were converted to timings of the mid-point of the egress feature so as to be consistent with previous authors, using a phase width of $0.0935$. We list also our determinations of the accretion disc radius $R_{disc}$ scaled by the binary separation $a$. This parameter is given for fifteen of the eighteen eclipses, omitting eclipses which are not complete enough for the disc radius to be determined with any precision.  Also listed in Table \ref{tab:wdparams} are the white dwarf fluxes in all three bands. These were calculated by measuring the height of the white dwarf egress feature in the model fits and then converted back to a magnitude scale following \citet{Smith02}. In Figure \ref{fig:egress} the $u'$-, $g'$- and $r'$-band lightcurves of the region around the white dwarf egress are plotted, with the accretion disc component of the emission subtracted from the data. This plot shows clearly that the height of the egress feature varies significantly from eclipse to eclipse (for example, compare egresses $15$ and $16$, taken on consecutive nights). This variability is apparent in all three bands, although in $r'$ the egress feature is very small and in $u'$ the signal-to-noise ratio is much lower, making it harder to determine the size of the egress accurately.

\begin{table*} 
\caption{White dwarf parameters obtained from MCMC fits to the individual lightcurves, as detailed in Section \ref{sec:entireeclipse}. We give the time of the white dwarf mid-egress $t_{mid-egress}$ and the white dwarf $R_{1}$ and accretion disc radii ($R_{1}$ and $R_{disc}$) scaled by the binary separation $a$, as obtained from the $g'$-band fits. $R_{disc} = R_{spot}$ in these model fits. We list also the $u'$-, $g'$- and $r'$- band magnitudes of the white dwarf, determined from the step height of the white dwarf egress feature.}
\label{tab:wdparams} 
\begin{tabular}{llll@{\,$\pm$\,}ll@{\,$\pm$\,}ll@{\,$\pm$\,}ll@{\,$\pm$\,}ll@{\,$\pm$\,}l} 
Eclipse  &cycle &\multicolumn{1}{c}{$t_{mid-egress}$}  &\multicolumn{2}{c}{$R_{1} / a$}  &\multicolumn{2}{c}{$R_{spot} / a $}  &\multicolumn{2}{c}{$u'$ mag}  &\multicolumn{2}{c}{$g'$ mag} &\multicolumn{2}{c}{$r'$ mag}\\
\hline		
$1$  &$-2200$ &$53246.145370(10)$   &$0.0037$ &$0.0003$     &$0.262$ &$0.002$          &\multicolumn{2}{c}{ }          &$18.83$ &$0.02$    &\multicolumn{2}{c}{ }\\
$2$  &$-2194$ &$53247.094620(12)$   &$0.0043$ &$0.0002$     &\multicolumn{2}{c}{ }                  &$19.08$ &$0.08$    &$18.53$ &$0.02$    &$19.57$ &$0.06$\\
$4$  &$0$     &$53594.198840(8)$    &$0.0031$ &$0.0008$     &$0.279$ &$0.002$          &$21.29$ &$0.18$    &$18.64$ &$0.01$    &$19.21$ &$0.02$\\
$5$  &$6$     &$53595.148038(5)$    &$0.0044$ &$0.0011$     &$0.269$ &$0.001$          &$19.50$ &$0.04$    &$18.51$ &$0.01$    &$19.19$ &$0.05$\\
$6$  &$18$    &$53597.046564(5)$    &$0.0052$ &$0.0008$     &$0.264$ &$0.002$          &$19.51$ &$0.05$    &$18.38$ &$0.01$    &$18.64$ &$0.01$\\
$7$  &$19$    &$53597.204771(6)$    &$0.0062$ &$0.0008$     &$0.271$ &$0.002$          &$19.14$ &$0.03$    &$18.62$ &$0.01$    &$18.89$ &$0.01$\\ 
$8$  &$24$    &$53597.995814(5)$    &$0.0060$ &$0.0017$     &$0.270$ &$0.001$          &$19.01$ &$0.03$    &$18.36$ &$0.01$    &$18.61$ &$0.01$\\
$9$  &$87$    &$53607.962749(5)$    &$0.0044$ &$0.0005$     &$0.268$ &$0.003$          &$18.73$ &$0.03$    &$18.29$ &$0.01$    &$18.54$ &$0.01$\\ 
$10$ &$88$    &$53608.120980(8)$    &$0.0040$ &$0.0006$     &\multicolumn{2}{c}{ }                  &$19.15$ &$0.06$    &$18.36$ &$0.01$    &$18.84$ &$0.02$\\ 
$11$ &$101$   &$53610.177655(4)$    &$0.0039$ &$0.0007$     &$0.253$ &$0.001$          &$19.63$ &$0.07$    &$18.74$ &$0.01$    &$18.23$ &$0.02$\\
$12$ &$106$   &$53610.968692(5)$    &$0.0043$ &$0.0005$     &$0.257$ &$0.001$          &$18.91$ &$0.04$    &$18.45$ &$0.01$    &$18.87$ &$0.01$\\ 
$13$ &$107$   &$53611.126905(5)$    &$0.0050$ &$0.0008$     &$0.254$ &$0.001$          &$18.47$ &$0.03$    &$18.20$ &$0.01$    &$18.41$ &$0.01$\\
$14$ &$112$   &$53611.917924(2)$    &$0.0056$ &$0.0010$     &$0.250$ &$0.002$          &$19.15$ &$0.17$    &$18.48$ &$0.02$    &$18.94$ &$0.05$\\
$15$ &$113$   &$53612.076134(4)$    &$0.0044$ &$0.0005$     &$0.258$ &$0.001$          &$18.64$ &$0.02$    &$18.25$ &$0.01$    &$18.65$ &$0.01$\\
$16$ &$119$   &$53613.025377(7)$    &$0.0043$ &$0.0006$     &$0.256$ &$0.001$          &$19.08$ &$0.05$    &$18.88$ &$0.01$    &$19.23$ &$0.02$\\
$17$ &$120$   &$53613.183571(7)$    &$0.0035$ &$0.0007$     &$0.260$ &$0.001$          &$19.57$ &$0.09$    &$19.07$ &$0.01$    &$19.56$ &$0.03$\\
$18$ &$133$   &$53615.240256(10)$   &$0.0054$ &$0.0010$     &$0.254$ &$0.002$          &$18.86$ &$0.07$    &$18.45$ &$0.01$    &$18.65$ &$0.02$\\
\hline
\end{tabular}
\end{table*}

\subsection{Eclipse timings}
\label{sec:timing}

We used the timings of the white dwarf mid-egress given in Table \ref{tab:wdparams} in order to calculate an updated orbital ephemeris for IP Peg. These data were complemented with eclipse timings given by previous authors (\citealt*{Goranskii85b}; \citealt{Wood89, Wolf93}). All times were converted to the barycentric dynamical timescale, corrected for light travel to the barycentre. A least-squares fit to all of these data yielded the ephemeris \\

\begin{figure}
\centering
\includegraphics[angle=270,width=0.48\textwidth]{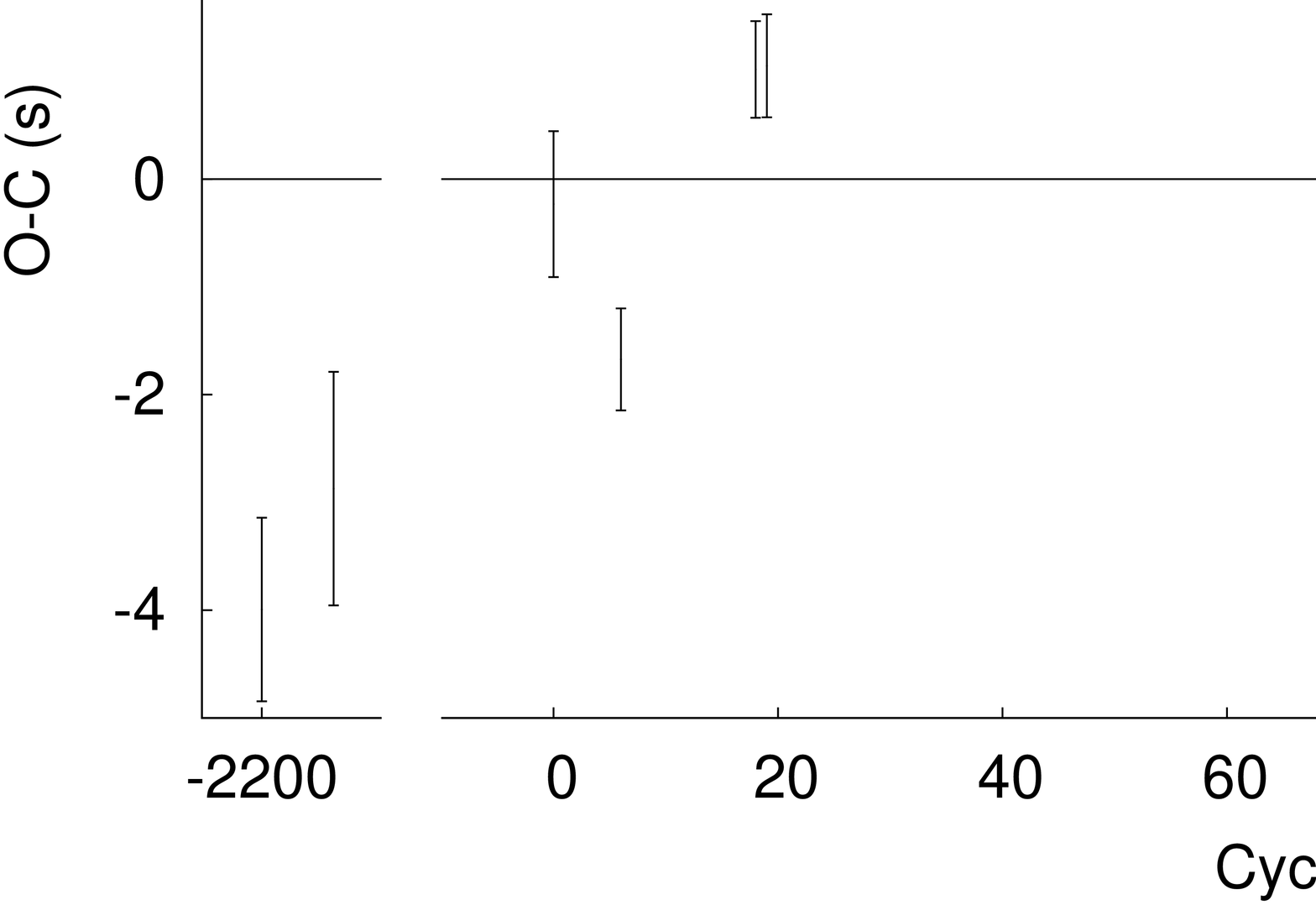}
\hfill
\caption{$(O-C)$ values plotted against cycle number, using the eclipse timings listed in Table \ref{tab:wdparams} and the linear ephemeris given in Section \ref{sec:timing}. We plot only our 2004/2005 datapoints, but used all available historical data in the determination of the ephemeris. Note the break in the x-axis of this plot between the 2004 and 2005 data.} \label{fig:ominusc} \end{figure}

$BMJD(TDB) = 53594.206270(1) +  0.1582061029(3)E$ \\ 

\noindent for the white dwarf mid-egress. This ephemeris fits well with our 2004/2005 data (Fig. \ref{fig:ominusc}), but the historical data still show the $\sim$$100$s deviation from a linear ephemeris discussed by \citet{Wolf93}, which they argued was caused by a third body in the system. An alternative explanation is that some of the historical timings are unreliable: the flux variations we observe in the egress feature (Figure \ref{fig:egress}) suggest that some observations with small telescopes may have failed to detect the egress feature at all. This is particularly likely in the instances when a very long egress was reported \citep{Wood86}, for which we see no evidence.

\section{DISCUSSION}
\label{sec:disc}

In this section we discuss the results of our model fits in more detail. The photometric method for parameter determination in eclipsing CVs is dependent on precise measurement of the contact phases of the white dwarf and bright spot eclipses. These measurements are used to constrain the geometry of the binary system, and derive the mass ratio and inclination. Our data affords the possibility of more precise determinations than previous authors, since we have for the first time identified the white dwarf ingress, and so can accurately determine the phase width of the white dwarf eclipse. We begin in Section \ref{sec:massratio} by discussing our determinations of the mass ratio and binary inclination, and the uncertainties on these measurements. We compare our results to determinations made both photometrically and spectroscopically by previous authors. In Section \ref{sec:binparam} we go on to investigate the individual components, and give masses, radii and radial velocity semi-amplitudes for both the white dwarf and the donor star. In Section \ref{sec:distance} we use these new parameter determinations to provide an updated estimate of the distance to IP Peg. In Section \ref{sec:wdcolours} we use our determinations of the white dwarf colours to determine its temperature, by comparing our measurements to theoretical cooling models. This task is complicated by the variability we observe in the white dwarf egress feature. In Section \ref{sec:photomodel} we detail a photoelectric absorption model which could account for this variability. As discussed in \citet{Townsley03}, the temperature of the white dwarf is driven by compressional heating of accreted matter, and so in Section \ref{sec:accretion} we discuss the implications of our temperature measurement for the accretion history of this system. Finally, in Sections \ref{sec:brightspot} and \ref{sec:accdisc} we examine the bright spot and accretion disc in more detail. In the case of the bright spot we determine its temperature and flux distribution, and for the accretion disc we examine changes in disc flux and radius over the course of our observations.

\subsection{Mass ratio and inclination}
\label{sec:massratio}

\begin{figure}
\centering
\includegraphics[angle=270,width=1.0\columnwidth]{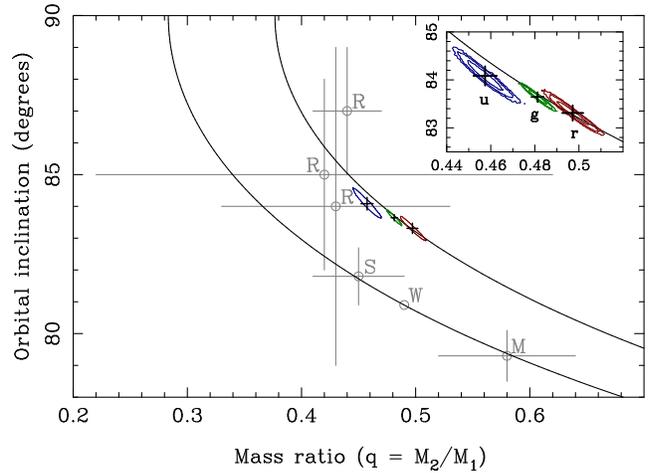}
\hfill
\caption{Mass ratio versus inclination, from our MCMC fits. We show the results of our fits to the phase-folded lightcurve, plotting the results in $u'$, $g'$ and $r'$ separately (blue, green and red, respectively). We plot points with the single parameter $1 \sigma$ errors and $95$\% joint confidence contours. The two solid lines running diagonally across the plot indicate the $q$/$i$ relationship for a phase width of $0.0863$ (bottom) and $0.0935$ (top). The inset shows an expanded view and the $68$, $95$ and $99$\% joint confidence contours. The grey points are measurements of $q$ and $i$ from the literature, and are labelled according to source as follows: $W$, \citet{Wood86}; $M$, \citet{Marsh88}; $S$, \citet{Smak02}; $R$, \citet{Ribeiro07}.} \label{fig:qvsi} \end{figure}

The phase width of the white dwarf eclipse is an observable quantity that is intrinsically linked to two physical properties: the mass ratio and the binary inclination. For a higher binary inclination the duration of the eclipse will be greater, thus to maintain the same phase width, as the inclination is increased, the size of the donor, and hence the mass ratio, must be decreased. There is therefore a unique relationship between these two properties \citep{Bailey79}. This degeneracy can be broken since we have an additional geometric constraint due to the ingress and egress of the bright spot. The path of the accretion stream and hence the position of the bright spot is modified by the mass ratio. With this additional information we can determine the mass ratio and inclination in this system. 

Previous studies of this system did not identify the white dwarf ingress feature, and so there has been some uncertainty in the phase width $\Delta\phi$ of the white dwarf eclipse. \citet{Wood86} suggested that the white dwarf ingress was blended with the bright spot ingress, implying $0.0863 < \Delta\phi < 0.0918$. We find the phase width to be $0.0935 \pm 0.0003$, implying either a higher mass ratio or a higher inclination than previously suggested, or some combination of the two. We find a much better fit for the high inclination case, due to the constraints implied by the bright spot.

Our MCMC results for the mass ratio versus inclination are plotted in Figure \ref{fig:qvsi}. The coloured contours and associated points show the single parameter $1 \sigma$ errors and $95$\% joint confidence contours. The other points plotted are various $q$ and $i$ determinations taken from the literature \citep{Wood86,Marsh88,Smak02,Ribeiro07}. The two black lines running across the plot show the unique $q$/$i$ relationship for phase widths of $0.0863$ and $0.0935$. \citet{Wood86} suggested $0.0863$ was the most likely value for the phase width: a number of authors have assumed this value and hence underestimated the inclination. 

If we now look at the values from our MCMC fits as plotted in Figure \ref{fig:qvsi}, it can be seen that the results from the different filters are not consistent at the $95$\% level. There are systematic uncertainties in our parameter determinations which dominate the uncertainties implied by the individual fits. We believe this to be due to the uncertainty in the measurement of the phase widths. While the position of the white dwarf egress is well defined, the white dwarf ingress, and to some extent, the bright spot ingress and egress, are still affected by flickering, even in the phase-folded lightcurve. In order to quantify the uncertainty in our system parameters, it was necessary to estimate the range of variation in the phase width as caused by the flickering. Our phase-folded lightcurve is the average of the eleven complete lightcurves. If the position of the white dwarf ingress is affected by flickering, it follows that if we choose a subset of these eleven to make a new phase-folded lightcurve, then we will measure a different phase width. It is likely that in some lightcurves the flickering influences the apparent position of the ingress more than in others, and so if the phase width is measured in enough of these subsets, the uncertainty in the phase width can be determined, from which a more realistic uncertainty for the physical parameters can be calculated.

\begin{figure*}
\centering
\includegraphics[angle=270,width=0.8\textwidth]{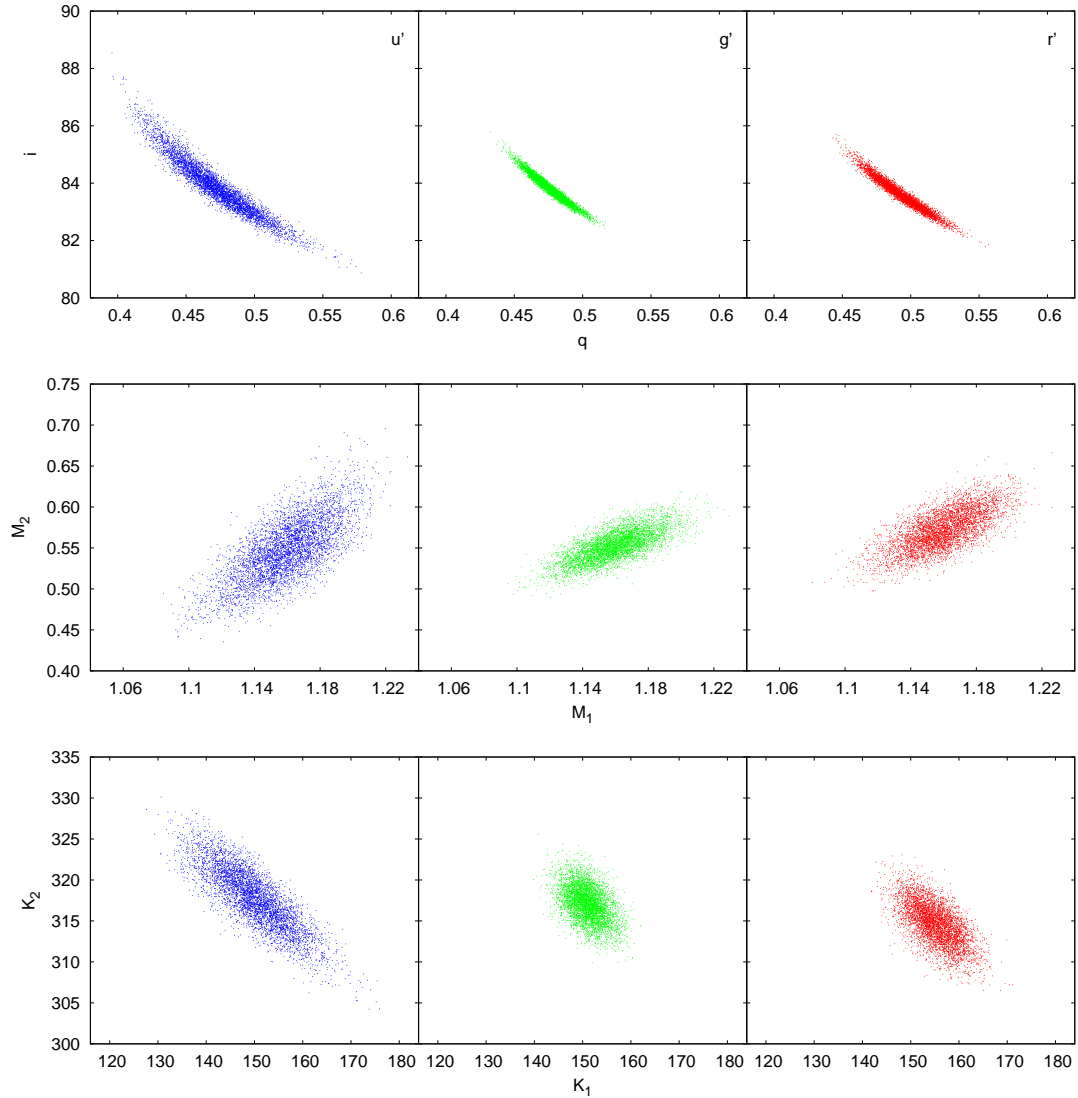}
\hfill
\caption{The results of our Monte Carlo simulation, which we use to determine the parameter uncertainties listed in Table \ref{tab:params}. We plot the mass ratio $q$ versus the inclination $i$, the white dwarf mass $M_1$ versus the donor mass $M_2$, and the white dwarf radial velocity semi-amplitude $K_1$ versus the donor semi-amplitude $K_2$. The three columns separate out the results from the individual bands (left to right, $u'$, $g'$ and $r'$).} \label{fig:montecarlo} \end{figure*}

We created a series of phase-folded lightcurves from different combinations of the complete lightcurves. Each phase-folded lightcurve was made up of eleven lightcurves (like the original), but  these eleven were selected using the bootstrap method \citep{Efron79,Efron93}. For each of these a minimisation was performed using the MCMC method, as before, and the contact phases of the ingress and egress features were computed. We additionally calculated the phase width of the white dwarf egress itself, using our previously determined $q$ and $i$ values, along with the weighted mean and variance of $R_1/a$ calculated from the values listed in Table \ref{tab:wdparams}. We subsequently calculated the covariance of the contact phases. These covariances are input into a Monte Carlo simulation through which we determined the uncertainty on the physical parameters of the system: $q$, $i$, the component masses $M_1$ and $M_2$, the binary separation $a$ and the radial velocity semi-amplitudes $K_1$ and $K_2$. One additional input into this calculation is an estimation of the extent to which the size of the white dwarf exceeds the Eggleton zero-temperature mass/radius relation (quoted in \citealt{Verbunt88}). For this the WD temperature estimate from Section \ref{sec:wdcolours} is used with the white dwarf cooling models of \citet*{Bergeron95} and \citet{Holberg06}. We find an oversize factor of $\sim$$1.04$ to be appropriate. The results of this simulation are plotted in Figure \ref{fig:montecarlo}, and all of these parameters are listed in Table \ref{tab:params}. Unlike the earlier MCMC fits, the uncertainties on these determinations are more realistic values which account for the systematic errors due to the uncertainty in the phase widths, and so the determinations are consistent in all three bands. The $g'$-band determinations are of course the most precise, due to the higher signal-to-noise ratio in these data.

We find the mass ratio in the $g'$-band data to be $q = 0.48 \pm 0.01$. This is consistent with, although more precise than, many of the previous determinations: the photometric studies of \citet{Wood89} for example suggested a value of $0.49$, and \citet{Smak02} reported $q = 0.45 \pm 0.04$. \citet{Wolf93} suggested a value of $0.6$ which is much higher than our finding, although previous photometric studies were limited by the difficulty in identifying the white dwarf ingress feature in the lightcurves. However, the spectroscopic study of \citet{Marsh88} also reported a higher value for $q$ ($0.58 \pm 0.06$).

The constraints on the binary inclination were reported by \citet{Wood86} to be $80.9 < i < 90$ deg, and previous authors have suggested values towards the low end of that range (e.g. \citealt{Smak02}, $i = 81.8 \pm 0.9$ deg). Our $g'$-band value of $83.81 \pm 0.45$ deg is higher than most previous determinations. As we remarked earlier, a confounding factor in previous studies has been the difficulty in identifying the white dwarf ingress feature, and the subsequent under-estimation of the phase width of the white dwarf's eclipse.

\subsection{The white dwarf and donor star parameters}
\label{sec:binparam}

\begin{table*} 
\caption{Binary parameters for IP Peg. We list the phase width of the white dwarf eclipse, the mass ratio $q$, the inclination $i$ and masses, radii and radial velocity semi-amplitudes of the two components. All of these parameters except the radii are determined from the Monte Carlo simulations described in Section \ref{sec:massratio}. The white dwarf radius is derived from the weighted mean of the values determined in the individual MCMC fits, and the donor radius is calculated using the approximation of \citet{Eggleton83}.}
\label{tab:params} 
\begin{tabular}{llr@{\,$\pm$\,}lr@{\,$\pm$\,}lr@{\,$\pm$\,}l} 
\hline
&&\multicolumn{2}{c}{$u'$-band}  &\multicolumn{2}{c}{$g'$-band}  &\multicolumn{2}{c}{$r'$-band}\\
\hline
\multicolumn{2}{c}{Phase width}     &$0.0933$ &$0.0006$         &$0.0935$ &$0.0003$     &$0.0941$ &$0.0003$\\
$q$ &		                        &$0.47$ &$0.03$             &$0.48$ &$0.01$         &$0.49$ &$0.02$\\
$i$ &(deg)	                        &$83.89$ &$0.96$            &$83.81$ &$0.45$        &$83.61$ &$0.53$\\	
$a$ &(\Rsun)	                    &$1.47$ &$0.02$             &$1.472$ &$0.009$       &$1.48$ &$0.01$\\
$M_1$ &(\Msun)	                    &$1.16$ &$0.02$             &$1.16$	&$0.02$         &$1.16$ &$0.02$\\
$R_1$ &(\Rsun)	                    &$0.0081$ &$0.0013$         &$0.0063$ &$0.0003$     &$0.0064$ &$0.0004$\\
$K_1$ &(km/s)	                    &$150$ &$7$                 &$151$ &$3$             &$155$ &$4$\\
$M_2$ &(\Msun)	                    &$0.55$ &$0.04$             &$0.55$	&$0.02$         &$0.57$ &$0.02$\\
$R_2$ &(\Rsun)	                    &$0.46$ &$0.02$             &$0.47$	&$0.01$         &$0.47$ &$0.01$\\
$K_2$ &(km/s)	                    &$318$ &$4$                 &$317$ &$2$             &$315$ &$3$\\
\hline
\end{tabular}
\end{table*}

In Table \ref{tab:params} we list our determinations of the masses, radii and radial velocity semi-amplitudes of the two components, as determined from our Monte Carlo simulation. The white dwarf radius $R_1$ is calculated from the weighted mean of the $R_1 / a$ values listed in Table \ref{tab:wdparams}, and the donor radius $R_2$ is calculated using the Roche-lobe volume radius approximation of \citet{Eggleton83}. 

If the $g'$-band results are used, the white dwarf mass is found to be $1.16 \pm 0.02$\Msun. This is larger than some previous determinations, for example \citet{Smak02} reported a value of $0.94 \pm 0.10$\Msun, but is in very close agreement with \citet{Watson03}, who report a value of $1.16$ -- $1.18$\Msun. We find the donor mass to be $0.55 \pm 0.02$ \Msun, which is greater than previous measurements. \citet{Smak02} reported a value of $0.42 \pm 0.08$ \Msun, and \citet{Watson03} found $M_2$ to be $0.5$\Msun. Our donor radius of $0.47 \pm 0.01$\Rsun \ is consistent with previous authors, but it is low given our mass determination, suggesting a donor that is slightly undersized compared to an equivalent main sequence star \citep{Baraffe98}. This is unusual for an accreting binary: donor stars tend to be oversized, due to either evolution or being out of thermal equilibrium as a result of mass loss. Our finding is consistent with a donor in thermal equilibrium. As we will show in Section \ref{sec:accretion}, we find the accretion rate to be much lower than would be expected for this system, and this might explain why the donor is not significantly perturbed. 
If the donor is in thermal equilibrium then this suggests the mass transfer rate has been low for a very long period of time, in excess of the thermal timescale ($\sim$$10^8$ yr, for a star with these parameters). Alternatively, the mass transfer might have begun very recently.

Recent work by \citet{Knigge06} has suggested a single, semi-empirical donor sequence for CVs with orbital periods $< 6$h. This sequence suggests a donor mass of $0.28$\Msun \ and radius of $0.38$\Rsun \ for a $P = 3.8$h CV. These values are much lower than our measurements; for our donor mass of $0.55$\Msun \ to be consistent with the sequence the binary period would need to be $\sim$$5.3$h. IP Peg is not the only system that fits poorly to the semi-empirical donor sequence of \citet{Knigge06}; we see in figure 9 of that paper that there are a number of other systems, such as RX~J0944, which have a donor that is of an earlier spectral type than would be expected for their binary period. Again, the low accretion rate we will report in Section \ref{sec:accretion} may imply an unusual accretion history for this system, which could account for this discrepancy.

Using our binary parameters, we calculated the radial velocity semi-amplitudes $K_1$ and $K_2$, and find them to be $151 \pm 3$ and $317 \pm 2$ km/s respectively. Our value of $K_2$ is larger than the long-established value ($298 \pm 8$km/s; \citealt*{Martin87}; \citealt{Martin89}), which was more recently supported by \citet{Beekman00}. However, the much better quality data of \citet{Watson03} give a value of $323.2 \pm 3.5$km/s. The consistency of our value derived purely from photometry with the best spectroscopic measurement is a rare test of the consistency of the photometric method (see also \citealt{Littlefair08} and \citealt*{Tulloch09}). The systematic uncertainty in spectroscopic determinations of $K_1$ is much higher than in $K_2$, and so previous values of $K_1$ in the literature as determined from the emission lines vary significantly ($175 \pm 15$ km/s, \citealt{Marsh88}; $118 \pm 10$ km/s, \citealt{Hessman89}). More recently, \citet{Smak02} used the three-body approximation to find a value for $K_1$ of $134 \pm 15$ km/s. This is consistent with our finding of $151 \pm 3$ km/s.

\subsection{Distance to IP Peg}
\label{sec:distance}

Using our new parameters for the donor star, we can update the estimation of the distance to IP Peg. The surface brightness method of distance determination \citep{Bailey81} uses the $K$-band magnitude to determine the distance to CVs. This method assumes all of the $K$-band flux originates in the donor star and is thus a lower limit, since the accretion disc will contribute an indeterminate amount of this flux (although this is small in the case of IP Peg). \citet{Szkody86} measured the $K$ flux and estimated the distance to be $130$ -- $142$pc. \citet{Bailey81} assumed a constant donor surface brightness in his relation, but subsequent authors have shown this quantity varies with the $V-K$ colour \citep{Ramseyer94,Beuermann06}. \citet{Froning99} reported an updated value of $121$pc using this refinement of \citet{Ramseyer94} and the $K$-band measurements of \citet{Szkody86}. We applied the \citet{Beuermann06} update of Bailey's method, and found the distance to IP Peg to be $151$pc. 
For this we used the \citet{Szkody86} $K$-band measurement, the donor radius listed in Table \ref{tab:params} and the same $V-K$ that was used by \citet{Froning99}, which is consistent with a donor spectral type of $\sim$M$4.8$. The uncertainty in our distance determination is likely to be dominated by the uncertainty in $V-K$. We took $\pm 1$ to be an appropriate estimate of the uncertainty in the spectral type, and determined the subsequent range in $V-K$ using Table 3 of \citet{Beuermann06}. We hence estimate the uncertainty in our distance determination of $151$pc to be $\pm 14$pc. The best trigonometric estimate of the distance is $152^{+44}_{-29}$pc (Thorstensen, private communication), which is consistent with our result.

\subsection{White dwarf colours and temperature}
\label{sec:wdcolours}

\begin{figure*}
\centering
\includegraphics[angle=270,width=1.0\textwidth]{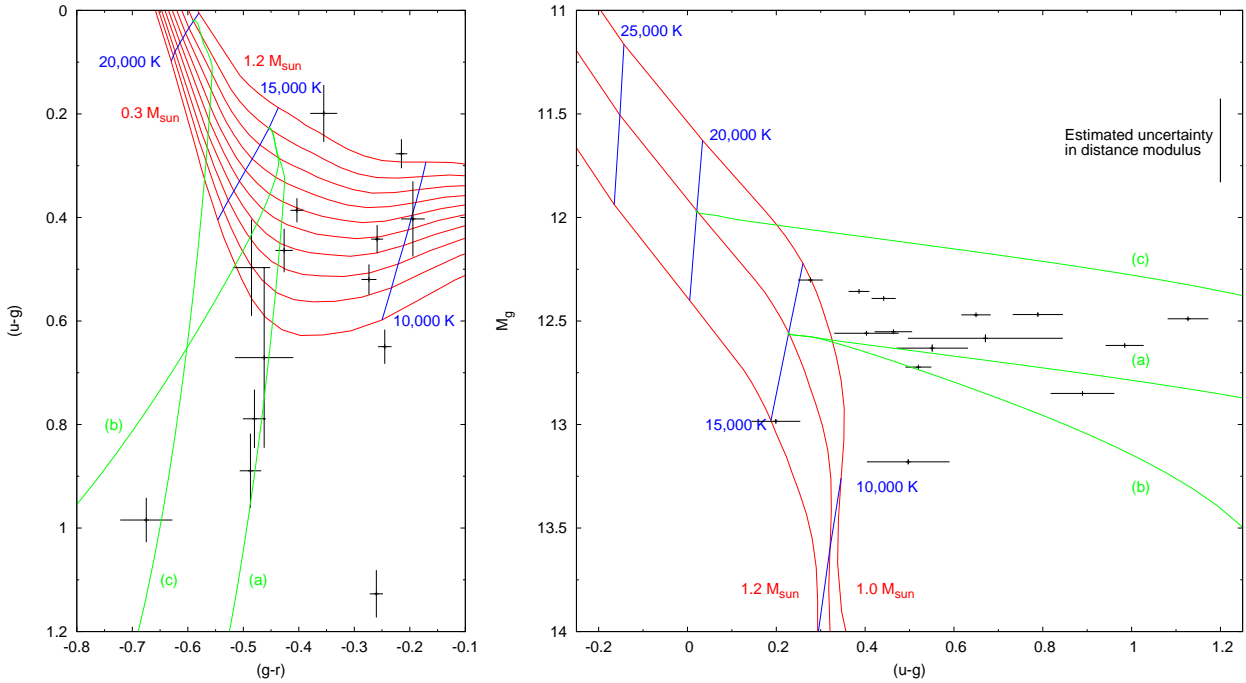}
\hfill
\caption{Using the eclipse determinations listed in Table \ref{tab:wdparams}, we plot datapoints showing the colours and magnitude of the white dwarf for each individual eclipse. In the left panel we plot the $u'-g'$ and $g'-r'$ colours for the white dwarf, and in the right panel we show the $u'-g'$ colour versus the absolute magnitude $M_g'$. For this we use our distance to IP Peg of $151$pc (Section \ref{sec:distance}), but we indicate to the right of the plot our estimate of the uncertainty in the distance modulus. The red lines that run from the top left area of each panel are the synthetic white dwarf tracks of \citet{Holberg06}. In the left panel we plot tracks for a white dwarf mass of $0.3$ to $1.2$\Msun, plotted every $0.1$\Msun, but for clarity in the right panel we just show the tracks for a mass of $1.0$, $1.1$ and $1.2$\Msun. Perpendicular to these tracks we plot isotherms (blue) every $5,000$K. The green lines are representative of the results of our opacity model, discussed in more detail in Section \ref{sec:photomodel}. Line $(a)$ assumes a white dwarf with a temperature of $15$,$000$K and a mass of $1.1$\Msun, and absorbing material with a temperature of $8$,$000$K and N$_H < 10^{23}$cm$^{-2}$. Line $(b)$ uses the same values as line $(a)$, except for the absorber temperature which is assumed to be $14$,$000$K. Line $(c)$ uses the same values as line $(a)$, except for the white dwarf temperature which is assumed to be $20$,$000$K.} \label{fig:wdcolour} \end{figure*}

The main motivation for this work was a photometric determination of the white dwarf temperature in this system. We see in Figure \ref{fig:egress} that, while the length of the white dwarf egress appears constant, its flux varies significantly on timescales of days. Using the white dwarf magnitudes listed in Table \ref{tab:wdparams} we calculate the $u'-g'$ and $g'-r'$ colours of the white dwarf for each individual eclipse. We plot these data in the left panel of Figure \ref{fig:wdcolour}. We plot also the synthetic white dwarf tracks of \citet{Holberg06} for a white dwarf mass range of $0.3$ to $1.2$\Msun.

This plot shows that as well as variation in all three bands, there are considerable variations in the white dwarf colours, particularly $u'-g'$. There is a suggestion of an inverse correlation between the $u'-g'$ and $g'-r'$ colours. This is due to large variation in the $u'$ and $r'$ fluxes and comparatively little change in the $g'$ flux, so things become a little clearer when we plot $u'-g'$ against the absolute $g'$-band magnitude $M_g'$ (Figure \ref{fig:wdcolour}, right panel). We assume a distance of $151$pc (Section \ref{sec:distance}), but plot the error bar corresponding to our estimate of the uncertainty in this determination. Despite this uncertainty, we see in Figure \ref{fig:wdcolour} a clear trend in the data. The datapoints with the lowest $u'-g'$ colours intersect with the synthetic tracks, and there are many other points which are much redder in colour. Not only are many of these points too red to correspond to a white dwarf of mass $\sim$$1.1$\Msun, they are too red to correspond to a white dwarf of any mass. 

We consider two possibilities for the variation in the white dwarf colours. Firstly, it is possible that there is some variable contamination from an additional temperature component, such as a band of hot, accreted material around the equator of the white dwarf \citep{Piro04}. If this was the case one expects the trend in the egress datapoints to be approximately in line with the direction of increasing white dwarf temperature in the \citet{Holberg06} tracks. However, what we actually see is a trend that is roughly perpendicular to the direction of increasing white dwarf temperature. The inverse correlation between the $u'-g'$ and $g'-r'$ colours is due to the fact that there is comparatively little change in the $g'$ flux compared to the (correlated) change in the $u'$ and $r'$ fluxes. It is difficult to reconcile this with a model that includes an additional temperature component. It is more likely that the variation in the white dwarf colours is due to photoelectric absorption, as a result of obscuration of the white dwarf by material from the accretion disc. This model can approximately explain the colour variations, and we examine this in more detail in Section \ref{sec:photomodel}. An occulting region of disc material could explain past reports of very extended white dwarf egresses in this system \citep{Wood86}. Alternatively these might have been failures to detect the variable egress at all. Given the short period of time over which we observed this system it is difficult to quantify the extent of this phenomenon.

We now attempt to estimate the white dwarf temperature in this system via comparison with the synthetic tracks. Figure \ref{fig:wdcolour} suggests a temperature of between $10,000$ to $15,000$K; it is difficult to be more precise than this due to the flux variations. Even so, this range is remarkably cool. Previous studies have suggested that a white dwarf in a CV with a period $> 3$h would have a typical temperature of between $25,000$ -- $35,000$K \citep{Winter03, Townsley03}.  Such a temperature is clearly inconsistent with our IP Peg data: this would require the white dwarf to be more than a magnitude more luminous in $g'$ and more than $1.5$ magnitudes brighter in $u'$. In order to plot the absolute magnitudes used in the right panel of Figure \ref{fig:wdcolour} we assumed a distance to IP Peg and the uncertainty in this measurement could allow a higher temperature, but judging from the parallax measurement this uncertainty is unlikely to be sufficient to reconcile a $25,000$K white dwarf with the observations. In addition, the $u'-g'$ / $g'-r'$ plot is independent of distance and also implies a temperature range of $10$,$000$ to $15,000$K for the white dwarf. 

In making this temperature determination we have assumed that our points with the bluest $u'-g'$ colour are indicative of a white dwarf that is largely unobscured. However, we also investigate the possibility that the accretion disc is opaque, blocking the bottom half of the white dwarf from view and so producing a situation where the obscuration of the white dwarf is at least $50\%$ at all times. With this modification we find the white dwarf temperature implied by our model fits is at most $\sim$$21,000$K. This is higher than our previous findings, but still significantly smaller than the predicted $25,000$ -- $35,000$K. Additionally, we will show in Section \ref{sec:photomodel} that the colour trends in the datapoints can be explained by an  absorption model if we assume a $15,000$K white dwarf, but not if we assume a temperature of $20,000$K or higher; hence we prefer the fits which imply a cooler white dwarf.

\subsection{An absorption model}
\label{sec:photomodel}

We propose that the variation in the white dwarf colours is due to occultation of the white dwarf by disc material, causing a significant and varying photoelectric absorption of the white dwarf flux. To test this we computed an opacity model and compared it to the data. Our intention here is not to make a precise and quantitative description of the absorption in this system, but to determine if a varying opacity is an adequate description for the flux variations we observe. 

We calculated the opacity for a range of absorber temperatures and neutral hydrogen column densities, allowing for hydrogen bound-free, free-free, bound-bound and H$^-$ opacities. A white dwarf spectrum was used to determine the source flux and we assumed LTE, an electron density of $10^{14}$ cm$^{-3}$, and we approximated the line broadening by assuming it to be Gaussian with an RMS of $50$ km/s. We convolved the resultant spectra with the SDSS filter profiles and hence calculated the change in the $u'$, $g'$ and $r'$ magnitudes for each absorber temperature and column density. These calculations give us the change in the source flux; we chose the unabsorbed colours to be those of a $15,000$K, $1.1$\Msun \ white dwarf, which is appropriate given our earlier measurements. The results of this model calculation are plotted in Figure \ref{fig:wdcolour} as tracks $(a)$ and $(b)$. These two tracks denote a constant absorber temperature of $8,000$ and $14,000$K respectively. The hydrogen column density increases along these tracks. If we compare the tracks to the white dwarf measurements we see that they lie approximately in line with the trend we observe in the datapoints, suggesting that absorption is an adequate explanation for the variations in white dwarf flux. There are some discrepancies; in particular a number of datapoints at low temperatures are not consistent with the tracks we plot, but we attribute this to the simplicity of our model. 

It was noted in Section \ref{sec:wdcolours} that it is possible that in all of our observations we see only a small fraction of the white dwarf flux, which would make our determination of the white dwarf temperature a significant underestimation. To test this we plot a third model track in Figure \ref{fig:wdcolour}, labelled $(c)$. This track uses the same parameters as track $(a)$ with the exception of the white dwarf temperature, which is assumed to be $20,000$K. We see this track is a poor fit to our observations, and so a white dwarf temperature of $20,000$K or greater is not supported by our opacity model. In addition, if the temperature of the white dwarf is greater than $\sim$$17,000$K, we find that there is no adjustment in the temperature of the absorbing material that can lead to the redwards movement in $g'-r'$ that would be required to fit the data.

Based on the neutral hydrogen column densities we use in our model, we calculate an N$_H$ of up to $\sim$$10^{23}$cm$^{-2}$ is necessary to account for the variation we observe in the white dwarf flux. For comparison \citet{Horne94} reported absorption of the white dwarf emission by disc material in the dwarf nova OY Carinae, but calculated an N$_H$ value that is an order of magnitude lower than we find for IP Peg. The higher N$_H$ value we find for IP Peg may be due to the very high inclination of this system. If this value is correct, it will be extremely difficult to detect the white dwarf in IP Peg at ultraviolet wavelengths. The ultraviolet continuum out-of-eclipse flux has been measured in IP Peg (\citealt{Hoard97}; \citealt*{Saito05}), and is consistent with a cool white dwarf and/or heavy absorption of the white dwarf flux.

\subsection{The long-term accretion rate}
\label{sec:accretion}

\citet{Townsley03} noted that the temperature of accreting white dwarfs is a good tracer of the medium term ($10^3$--$10^5$ yr) accretion rate. Our value for the IP Peg white dwarf implies a mean accretion rate of $< 5 \times 10^{-11} M_{\odot}$ yr$^{-1}$. The expected rate for a dwarf nova of this period is $\sim$$2 \times 10^{-9} M_{\odot}$ yr$^{-1}$: more than $40$ times greater. The value we find for the mean accretion rate is much closer to the instantaneous rate, which we estimate to be $\sim$$1.8 \times 10^{-10} M_{\odot}$ yr$^{-1}$ by following the calculation of \citet{Marsh88} and substituting in our new distance determination.

Our determination of the mean accretion rate implies a mass transfer timescale of $\sim$$10^{10}$ yr, much greater than the thermal timescale for the donor, which is $\sim$$10^8$ yr. We would therefore expect the donor in IP Peg to be unperturbed by the mass transfer, in contrast to the donors in many other CVs which are observed to be oversized when compared to isolated main sequence stars of equivalent mass. As we noted in Section \ref{sec:binparam}, we do find a mass and radius for the donor that are consistent with those of a main sequence star in thermal equilibrium.

The low temperature of this system suggests there could be a selection bias in the sample of long-period systems presented in \citet{Townsley03}, and further detailed studies of eclipsing systems are necessary in order to determine the extent of this bias. At this stage, for example, it is possible that mass transfer in IP Peg has begun very recently and so the temperature in this system is a poor indicator of the long term rate. One possible mechanism is the `hibernation scenario' proposed by \citet{Shara86}, in which, following a classical nova eruption and a white dwarf cool-down period, the binary is detached for $10^3$ -- $10^6$ yr. \citet{Shara86} argue that an old nova emerging from this period of detachment (due to angular momentum loss bringing the donor back into contact with its Roche lobe) might first appear as a dwarf nova. 

Alternatively, we can take our low value for the mass transfer rate over the past $10^3$--$10^4$ yr in this system to be correct. This is surprising, given the contention that long-period dwarf novae have a high mean \Mdot \ despite their observed instantaneous values. The low temperature of IP Peg would suggest it has sustained an accretion rate well below the expected mean rate for more than $1000$ years. If all `steady-state' CVs go through low accretion rate dwarf nova phases, then this result implies the length of these phases can be very long indeed. 

The third possibility is that there are problems with the classical picture of CV evolution as driven by angular momentum loss. The orbital period distribution of CVs \citep{Spruit83} has traditionally been explained by the disrupted magnetic braking model \citep{Robinson81}, whereby angular momentum loss above the $2$ -- $3$ h `period gap' is due to a combination of gravitational
radiation and magnetic stellar wind braking, and below the gap is due to gravitational radiation only. It follows that angular momentum loss is higher at longer periods, and since angular momentum loss is proportional to \Mdot,  \Mdot \ is expected to be high at long periods. The existence of low \Mdot \ dwarf novae like IP Peg at long periods conflicts with this picture. Some authors suggest alternatives, such as  \citet{Willems07}, who consider the effect of circumbinary discs on CV evolution. The theoretical orbital period distribution of \citet{Willems07} is a reasonable reproduction of the observed distribution, and they find that systems with unevolved donors need not detach and evolve below the period gap as in the disrupted magnetic braking model. In their model different CVs can have very different long-term \Mdot s, and some systems can exist as dwarf novae throughout their lifetime. This would explain the \Mdot \ we measure in IP Peg, although study of further systems is necessary to distinguish between this and the other possibilities.

\subsection{The bright spot}
\label{sec:brightspot}

\begin{figure}
\centering
\includegraphics[angle=270,width=1.0\columnwidth]{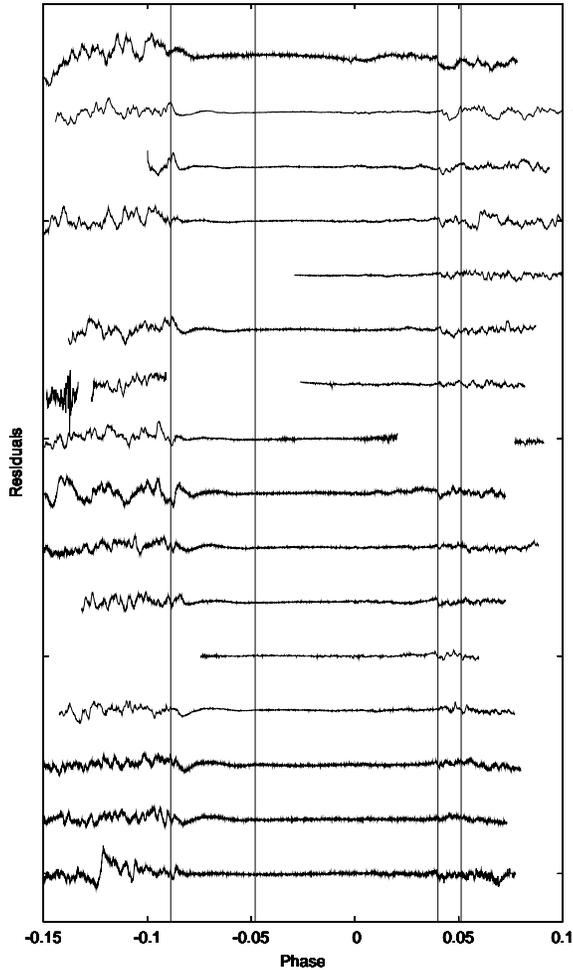}
\hfill
\caption{The residuals for the model fits to the individual $g'$-band lightcurves. The four vertical lines show (from left to right) the beginning and end points of the bright spot ingress, followed by the begining and end points of the bright spot egress. A phase of zero corresponds to the mid-point of the white dwarf egress. We omit eclipse $2$ because it includes no data for the bright spot, and eclipse $3$ because the flare event near the white dwarf egress precludes a precise fit.} \label{fig:brightspot} \end{figure}

It is apparent from the lightcurves (Figure \ref{fig:lightcurves_1}) that most of the flickering originates in the bright spot. The bright spot is modelled as a linear region which approximately follows the outer circumference of the disc, with a luminosity which decreases approximately exponentially from the point of impact of the accretion stream (see Appendix \ref{sec:appendix}). The geometry is such that it is the impact region which is eclipsed first, and is first to emerge from eclipse. In Figure \ref{fig:brightspot} we plot the residuals from the model fits to the individual lightcurves, and we indicate on this figure the beginning and end points of the bright spot ingress and egress features. It is apparent in this figure that the majority of the variation due to flickering is localised in the impact region, since we see the residuals very quickly flatten off following the beginning of the ingress, but return immediately following the beginning of the egress.

\begin{figure*}
\centering
\includegraphics[angle=270,width=1.11\columnwidth]{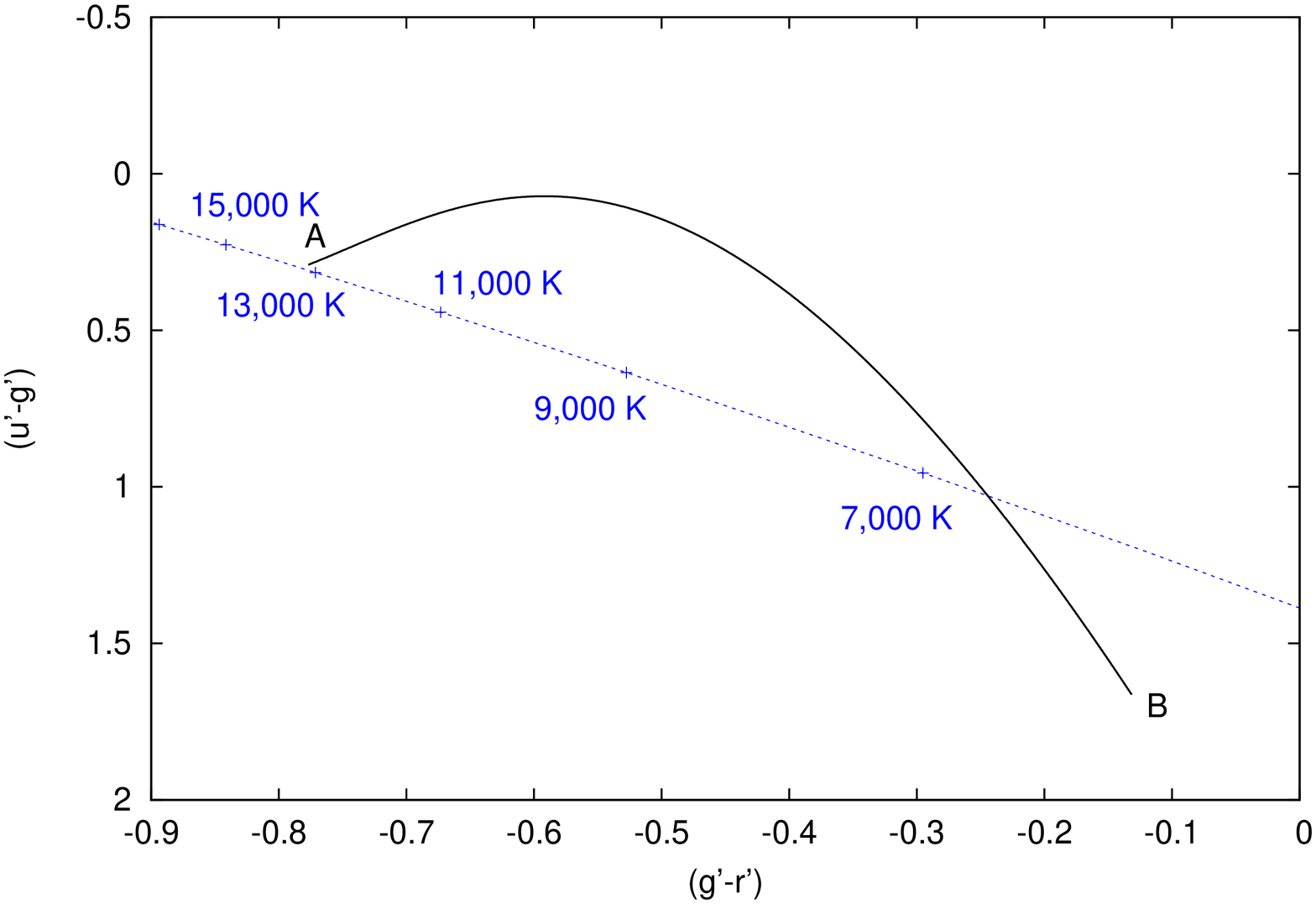}
\hspace{0.5cm}
\includegraphics[angle=270,width=0.75\columnwidth]{figures/brightspot_diag.ps}
\caption{Left: The solid line (black) shows the colour variation along the bright spot, from the point of impact of the gas stream (labelled `A'), to a point $10$ scale lengths away from this impact region (labelled `B'). The dashed line (blue) shows for comparison the blackbody colours for a temperature ranging from $\sim$$6,000$ -- $17,000$K. Right: Schematic of the primary Roche lobe. The dashed line shows the extent of the Roche lobe, the dotted line shows the accretion disc radius and the central point is the white dwarf (all to scale). The line originating at the L1 point is the gas stream, and the bright spot is plotted from the impact region `A' to `B' as a series of circles, with the area of each circle representing the surface brightness at that point.} \label{fig:bscolour} \end{figure*}

In Figure \ref{fig:bscolour} we plot the colour variation along the bright spot. For this we used the bright spot parameters listed in Table \ref{tab:mcmcresults} with Equation \ref{eq:bs} to calculate analytically the flux in each filter at each point along the length of the spot. We plot the bright spot colour from the point of impact of the gas stream on the accretion disc (labelled `A' in Figure \ref{fig:bscolour}) to a point $10$ scale lengths ($l$; see Appendix \ref{sec:app_bs}) away from the impact region (labelled `B'). We plot also a line showing the colours of blackbodies with temperatures in the $\sim$$6,000$ -- $17,000$K range. At the point of impact the bright spot is close in colour to a $13,000$K blackbody, and as we move away from the impact region the colours of the bright spot decrease approximately in the direction of decreasing blackbody temperature. The spectroscopic study of \citet{Marsh88} also found the closest blackbody fit to the bright spot was for a temperature of $13,000$K. The shape of the bright spot line is not very important since it is very model dependent and as we note in Appendix \ref{sec:app_bs}, our bright spot model does not describe the IP Peg bright spot perfectly. However, the fact that the general trend of the line is in the direction of decreasing blackbody temperature is clear evidence for cooling along the line of bright spot elements. The variable temperature of the bright spot in IP Peg will cause it to vary according to the wavelength of observation. For instance near-infrared data suggest a cooler $10,000$K \citep{Froning99}, and hotter parts than we see may dominate in the far-ultraviolet.

The points `A' and `B' are also labelled in the schematic of the primary Roche lobe. An additional point to note from the schematic is that the angle of the bright spot line of elements is such that it runs approximately along the circumference of the accretion disc. This behaviour is not enforced in our model, but seems physically appropriate: the incoming material from the gas stream is thermalised at point `A' when it strikes the dense accretion disc, and as it cools, it propagates around the disc along a line of approximately constant radius and disc density.

\subsection{The accretion disc}
\label{sec:accdisc}

\subsubsection{Accretion disc radius}
\label{sec:discrad}
\begin{figure}
\centering
\includegraphics[angle=270,width=1.0\columnwidth]{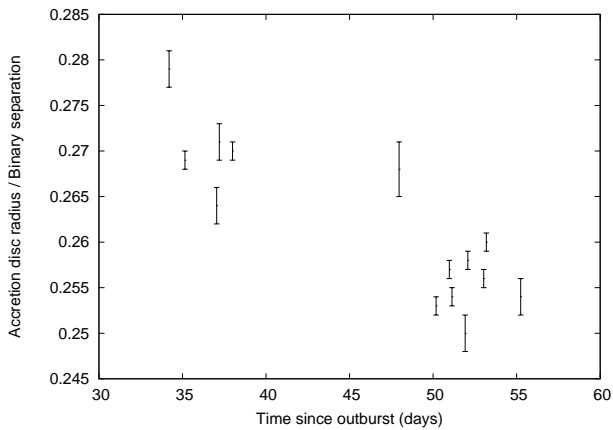}
\hfill
\caption{Variations in accretion disc radius over the course of the August/September 2005 observations. We plot the accretion disc radius $R_{disc}$ scaled by the binary separation against the MJD time since the last outburst.} \label{fig:discrad} \end{figure}

Data collected by the American Association of Variable Star Observers\footnote{http://www.aavso.org/} shows that IP Peg was observed to go into outburst approximately $35$ days prior to the beginning of our August 2005 observations. In Figure \ref{fig:discrad} we plot the accretion disc radius scaled by the binary separation against the MJD time since the detection of the outburst, using the $g'$-band model fits. We see there is a decline in disc radius over the course of the $\sim$$20$ days covered by our observations. During outburst the accretion disc radius increases as angular momentum is transported outwards due to the disc viscosity, and following the outburst the disc radius begins to shrink as low angular momentum
material is added to the disc via the gas stream. It is this decline we observe in Figure \ref{fig:discrad}. We compared these values with previous determinations of the disc radius given by \citet{Wood89} and \citet{Wolf93}. In Figure 6 of \citet{Wolf93} disc radius is plotted as a function of time since outburst, and our measurements are consistent with the rate of decline shown in their figure.

\subsubsection{Accretion disc flux}

The current and well-established paradigm for outbursts in dwarf novae is the thermal-viscous accretion disc instability model \citep{Osaki74}. This model proposes that the outbursts are due to the bi-stable nature of accretion discs when $T_{disc} \sim$$10,000$K, at which point hydrogen in the disc changes from a neutral to an ionised state \citep{Meyer81}. During quiescence, matter is stored in the disc and at a critical amount the instability sets in and a significant fraction of this mass is very rapidly accreted onto the central object. There are a number of detailed one-dimensional calculations of the disc instability model (see, e.g., \citealt{Hameury98}; \citealt{BuatMenard01}), and among their predictions is a significant increase in the disc luminosity during quiescence, since as matter is stored in the disc the surface density increases, prompting an increase in the effective temperature. However this is not observed: long term studies of dwarf novae show the visible flux is approximately constant during quiescence \citep{Cannizzo92}. Much of the existing data has been collected by amateurs and is of variable quality, but the models predict an increase of $\sim$$1$ --$3$ visual magnitudes, which should be easily detectable in these data \citep{Smak00}. It has been proposed that these observations could be explained by a hot, high-viscosity inner region of the disc which slowly cools during quiescence, and the decrease in flux from this region counterbalances the increase in the effective temperature of the outer disc \citep*{Truss04}. 

\begin{figure}
\centering
\includegraphics[angle=270,width=1.0\columnwidth]{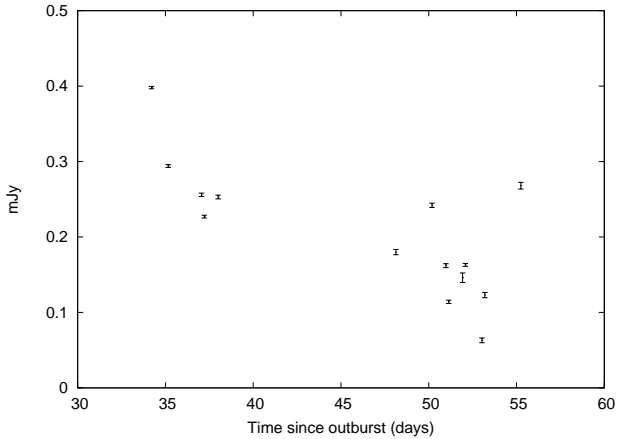}
\hfill
\caption{Variations in accretion disc flux over the course of the August/September 2005 observations. We plot the $g'$-band accretion disc flux at the point of the white dwarf egress against the MJD time since the last outburst.} \label{fig:discflux} \end{figure}

Eclipsing systems such as IP Peg allow for a more precise measurement of disc flux variations, since during the phase range over which both the white dwarf and the bright spot are eclipsed, the only contributors to the observed flux are the donor star and the accretion disc. The donor star component should be essentially constant, and so by measuring the average flux at a constant phase for every eclipse, we can determine the luminosity evolution of the accretion disc. We chose to measure the flux at the phase of the white dwarf egress, since at this point we observe a complete half of the disc and so our results should be unbiased by temporal radial variations. We used our model fits to subtract the white dwarf contribution at this point. We plot the average flux over a $10$s period at mid-egress in Figure \ref{fig:discflux}, and see an apparent decrease in disc flux over the period covered, in contrast to the increase which is predicted by the disc instability model. Our findings are consistent with the `mirror eclipse' phenomenon reported by \citet{Littlefair01}, which they interpreted as eclipses of the donor star by the optically thin outer edge of the accretion disc. This conflicts with the disc instability model, which requires an optically thick disc for physically viable values of the disc viscosity parameter. Note also that the decrease in brightness we observe (from $\sim$$0.4$ to $0.1$mJy) cannot be explained purely in terms of the disc radius variations we discussed in Section \ref{sec:discrad} , since we see in Figure \ref{fig:discrad} that the change in disc area over the course of our observations is only $\sim$$25\%$. Our results demonstrate a genuine dimming of the disc over the course of our observations. We also do not observe any evidence for the hot inner disc region proposed by \citet{Truss04}.

\section{CONCLUSIONS}
\label{sec:conc}
In 2004 and 2005 we took eighteen separate observations of the eclipse in the $3.8$h period dwarf nova IP Peg. We used the triple-band camera ULTRACAM mounted on the William Herschel Telescope, and observed simultaneously in the Sloan $u'$-, $g'$- and $r'$-bands. By phase-folding these data we identify for the first time the white dwarf ingress in this system. The phase width of the white dwarf eclipse is $0.0935 \pm 0.0003$, significantly higher than the standard value used by previous authors. This new value implies a higher orbital inclination or a higher mass ratio or some combination of the two. Our model fits strongly favour a high orbital inclination, which we find to be $83.81 \pm 0.45$ deg. Our determinations of the masses of the two components are higher than those reported by  previous authors.

We find the brightness of the white dwarf egress feature to be highly variable in flux and colour. We suggest this is due to photoelectric absorption as a result of obscuration of the white dwarf by material from the accretion disc. We find the duration of the egress feature to be consistent in length, and so suggest that past reports of very extended egresses were due to failures to detect the egress at all, as a result of this varying obscuration.

It is difficult to determine an accurate temperature for the white dwarf due to the obscuration, but by comparing our data with synthetic white dwarf models we find the most likely temperature to be in the range of $10,000$ - $15,000$K. This is very low for a dwarf nova above the period gap and, following \citet{Townsley03}, implies a mean accretion rate over the past $1,000$ - $10,000$ years of $< 5 \times 10^{-11} M_{\odot}$ yr$^{-1}$. This is more than $40$ times lower than the expected rate. Additionally we find a mass and radius for the donor that are consistent with those of a main sequence star in thermal equilibrium, which also suggests a very low accretion rate. Unless the mass transfer in IP Peg has begun very recently, these findings imply either that CVs can sustain accretion rates well below the expected rate for very long periods of time, or that the classical picture whereby the long-term accretion rate scales with period is not correct.

\section*{ACKNOWLEDGEMENTS}
CMC and TRM are supported under grant ST/F002599/1 from the Science and Technology Facilities Council (STFC). ULTRACAM and SPL are supported by STFC grants PP/D002370/1 and PP/E001777/1. The results presented in this paper are based on observations made with the William Herschel Telescope operated on the island of La Palma by the Isaac Newton Group in the Spanish Observatorio del Roque de los Muchachos of the Institutio de Astrofisica de Canarias.  We also acknowledge with thanks the variable star observations from the AAVSO International Database contributed by observers worldwide and used in this research. This research has made use of NASA's Astrophysics Data System Bibliographic Services and the SIMBAD data base, operated at CDS, Strasbourg, France. We thank Chris Watson for providing access to his spectroscopic data, and John Thorstensen for preliminary information on the parallax of IP Peg. We would additionally like to thank the referee, Robert C. Smith, for comments which led to a number of improvements to this paper.

\bibliography{ippeg}

\appendix

\section{The light-curve modelling code}
\label{sec:appendix}

In this section we provide a concise description of the code, {\sevensize LCURVE}, that we
wrote in order to model the light-curves of this paper; the same code has been
used in \citet{Pyrzas09} and \citet{Southworth09}. The aim of the code was
to provide a flexible framework for fitting light-curves characteristic of
eclipsing dwarf novae and detached white dwarf / M dwarf binary stars. {\sevensize LCURVE} 
was developed as a generalisation of the code used in \citet{Horne94}
to model the dwarf nova OY~Car and subsequently employed in the analysis of 
ULTRACAM photometry of several other systems \citep{Feline04,Feline05,Littlefair06}. 

The binary is defined by up to four components: white dwarf, secondary star,
accretion disc and bright-spot. In the case of detached systems, the latter
two are disabled. Each component is described by a set of small, flat
elements, each of specified area, position, orientation and brightness. The
light-curves are calculated by summing the contributions from all faces which
are oriented towards the observer, allowing for eclipses by the two
stars. Additionally, it is possible to make the accretion disc opaque or
transparent. The contribution of a face depends upon how it is oriented with
respect to the line of sight. Denoting the angle between the perpendicular to
a face and the vector pointing towards Earth as $\theta$, there is no
contribution if $\theta > \pi/2$ radians, or equivalently if $\mu = \cos \theta
< 0$. Otherwise the contribution is typically specified using a limb-darkened
Lambert's law behaviour:
\begin{equation}
 I \propto \mu \left(1 - \epsilon + \epsilon \mu\right),
\end{equation}
where $\epsilon$ is a linear limb-darkening coefficient (the code also allows
for quadratic limb-darkening).

The effect of finite length exposures is allowed for by calculating multiple
points covering an exposure and using trapezoidally-weighted averaging. Thus
$30$ second-long exposures of a white dwarf eclipse can in the model be split
into $7$ points separated by $5$ seconds for example.

In \citet{Horne94}, the overall scaling of the various components was
handled using singular-value decomposition (SVD), allowable if all components
contribute linearly to the light-curve. The advantage of this approach is
speed: during the fitting process, several potential free-parameters are
immediately removed at an early stage, allowing more time for optimisation of
the remaining model parameters. However, it is not always a valid approach, in
particular if ``reflection'' (more properly reprocessing) is significant, then
one cannot assume a simple linear combination, since doubling the
contribution of the white dwarf, say, may also imply an altered contribution
from the heated secondary star. Thus in {\sevensize LCURVE} one can switch off the SVD
scaling, falling back on the physical model to handle the effects of
reflection.

We experimented with several forms of minimisation: the simplex method
implemented via ``amoeba'' \citep{Press02}, Powell's method and the
Levenburg-Marquardt method. The simplex method has the advantage of
robustness, while Levenburg-Marquardt can deliver uncertainty estimates once
it has reached a minimum. However we found neither to be particularly secure
in reaching the minimum $\chi^2$ for the rather complex fits that we undertook
in this paper. Early on in the fitting we realised on several occasions that
we were trapped in local-minima, but even when we were not, the routines would
often report that the fits were converged when subsequent work revealed this
not to be the case. We believe that this was largely caused by degeneracies in
the model which are very hard to remove entirely. There is for instance a
fundamental degeneracy between mass ratio $q$ and orbital inclination $i$ that
cannot be avoided. Ultimately, the method that helped above all in obtaining
the final fits, and which, after extensive use, gave us confidence that we did
achieve global minima, was the Markov Chain Monte Carlo (MCMC) method. Our
standard technique evolved to become (i) carry out an initial minimisation
with the simplex method, and (ii) then run an MCMC chain until it reaches an
essentially flat run of $\chi^2$ with fit number, consistent with the
statistical fluctuations characteristic of the MCMC method. For our main fits
we performed the same process from 10 independent starting models to check
that we obtained consistent results. We typically required chains of
$10$,$000$ fits or more to reach the desired quasi-static state. We strongly
recommend this technique in all similar work to guard against premature claims
of ``convergence''.

Apart from the general physical parameters mentioned above and others
associated with the individual components, the other parameters needed to
define the model, were computational, such as the number of elements needed to
define the various components. Our strategy with these was simply to make them
large enough that they had no significant effect upon the best-fit parameters.
We finish by describing the model used for the individual components in more
detail.

\subsection{The white dwarf}
We modelled the white dwarf as a linearly limb-darkened sphere. Although we
also enabled the possibility of Roche-distortion, we never used it in the fits
presented here. The eclipse of the three other components by the white dwarf
was included by default, although its effect is tiny in the case of IP~Peg
since the white dwarf is so small.

\subsection{The secondary star}
The secondary star was modelled with full accounting for distortion in Roche
geometry. Again linear limb-darkening was assumed, and heating by flux from
the white dwarf was included along with gravity darkening. None of this is
especially significant for IP~Peg as the secondary star contributes relatively
little. The important feature is of course the eclipse, again calculated
accounting for Roche geometry.

\subsection{The accretion disc}
We model the accretion disc as a symmetric, flattened disc with user-defined
inner and outer radii. In the case of IP~Peg we forced the inner radius to
match the radius of the white dwarf. When fitting the individual lightcurves the outer disc radius $R_{disc}$ was set to match the radius of the bright-spot ($R_{spot}$, next section) although in the case of the phase-folded lightcurves we found a small improvement in the fit when we allowed $R_{disc}$ to be a free parameter. These fits show $R_{spot}$ to be somewhat smaller than $R_{disc}$, particularly in the $u'$-band. This may in part be due to the disc being non-circular, but note also $R_{disc}$ is rather poorly constrained and the uncertainties in these values are likely much greater than the formal errors listed in Table \ref{tab:mcmcresults}.

In between the user-defined
inner and outer radii, the
height of the disc was defined to follow a power-law in radius, i.e.
\begin{equation} 
h(R) = h_0 R^\alpha,
\end{equation}
where $h_0$ and $\alpha$ are parameters of the model. The height parameter
$h_0$ and the radius $R$, along with all other lengths, were scaled in terms
of the binary separation, $a$. We fixed $h_0 = 0.02$ and $\alpha = 1.5$. Since
the outer disc radius $\approx 0.3 a$, this is equivalent to a very thin disc.
The feeble emission from the disc meant that $h_0$ and $\alpha$ have little
overall impact, and so they were not optimised during the fits. The surface
brightness of the disc was also a power-law in radius, with the overall level
defined by the temperature at the outermost radius of the disc, translated
into a surface brightness assuming a blackbody spectrum given the central
wavelength of the filter concerned. It is important to realise that this does not
mean that we are saying that the radiation from the disc is that of a
blackbody; it is just one way of normalising the overall contribution from the
disc. Nevertheless, our model can certainly be criticised for the assumption
of a power-law variation in surface brightness and for its symmetry. Neither
is necessarily correct, on the other hand there is nothing in the data that
definitely points the way to a better alternative. We did allow the power-law
exponent of the surface brightness to be optimised, since it has no clear 
value \emph{ab initio} for quiescent discs, and since it has a significant
effect, particularly during the egress phases of the eclipse.

\subsection{The bright-spot}
\label{sec:app_bs}
The bright-spot was the hardest component to model, and indeed, as Figure \ref{fig:phasefold}
shows, we never achieved an entirely satisfactory fit. The bright-spot is of
course a uniquely strong feature of IP~Peg, and so it is to be expected that
it would cause difficulties. We model the bright-spot as a series of elements
which lie along a straight line in the orbital plane (as we illustrate in the Figure \ref{fig:bscolour} schematic). The angle ($\phi$) that the line
makes to the line of centres between the two stars is a key defining feature
of the bright-spot since it is this, in conjunction with the distance of the
bright-spot from the white dwarf and the mass ratio and inclination, which
determines the relative lengths of the bright-spot ingress and egress
features, which are both well-resolved in our data. Each element was then
assumed to contribute some fraction ($f_c$) that was constant in phase together with
some fraction of light modulated according to the angle of the element. In our
model the elements were assumed oriented perpendicular to the orbital plane
(although we can allow them to have arbitary orientation), so they give
light-curves that peak when we see them face-on. The angle $\psi$ is defined as the angle away from the perpendicular to the line of bright spot elements at which the light from the bright spot is beamed. In this manner, the variation
of the orbital ``hump'' and the relative size of the bright-spot ingress and
egress features can be fitted. Note that due to the way they are defined there is a strong correlation between the angles $\phi$ and $\psi$. This degeneracy can be broken if the observed phase range covers the entire orbital ``hump'', but this is not the case for our IP Peg data. These angles are therefore poorly constrained by our model and we see in Table \ref{tab:mcmcresults} that the fitted values for $\phi$ and $\psi$ vary significantly between the different bands. Note however that $\phi + \psi$ is consistent, and so the direction of peak emission is the same for all photometric bands. Finally, we should point out that we find no correlation between these parameters and $q$ or $i$, and so the fact that these bright spot angles are poorly constrained does not introduce any additional uncertainty in the physical parameters listed in Table \ref{tab:params}.

 The surface brightness of the elements in the
bright-spot were specified by
\begin{equation}\label{eq:bs}
 S \propto \left(\frac{x}{l}\right)^\beta \exp \left[-
\left(\frac{x}{l}\right)^\gamma\right],
\end{equation}
where $x$ is the distance along the line defining the bright-spot,
$\beta$ and $\gamma$ are two more power-law exponents which allow some flexibility
in how the brightness varies, and $l$ is a scale-length. The final element
of the bright-spot is its position within the binary. The maximum of the
surface brightness occurs at 
\begin{equation}\label{eq:head}
 x = l \left(\frac{\beta}{\gamma}\right)^{1/\gamma}. 
\end{equation}
We define this location to lie upon the ballistic gas-stream, at a 
specified distance from the white dwarf $R_{spot}$, this ``bright-spot radius'' being the
final parameter needed to define the bright-spot. The path of the gas stream
is calculated by integration of the equations of \citet{Lubow75}.

Compared to the modelling of the bright-spot in \citet{Horne94} and the
related papers mentioned before, the parameter $\gamma$ allows increased
flexibility in the specification of the bright-spot, since in the earlier
work, $\gamma = 1$ by default. This extra flexibility comes at the cost of
increased degeneracy; the MCMC method was essential to cope with this. Despite
this, as mentioned earlier, the strong bright-spot of IP~Peg is still not
modelled perfectly by our code, but we felt that adding yet more complexity to
the model would not be justified by any increased confidence in the
fundamental parameters such as $q$ and $i$.

\begin{table} 
\caption{Results from our MCMC fits to the phase-folded lightcurves, as detailed in Section \ref{sec:phasefold}. $q$: the mass ratio. $i$: the binary inclination. 
$R_1$: the white dwarf radius. $R_{disc}$: the accretion disc radius. $R_{spot}$: the distance between the maximum of the
surface brightness of the bright spot (Equation \ref{eq:head}) and the white dwarf. $a$: the binary separation. $\beta$ and $\gamma$: the power-law exponents for the bright spot. $\delta$: the exponent of surface brightness over the accretion disc. $l$: the bright spot scale-length. $f_c$: the fraction of the bright spot taken to be equally visible at all phases. $\phi$: the angle made by the line of elements of the bright spot, measured in the direction of binary motion from the line from accretor to donor. $\psi$: the angle away from the perpendicular to the line of elements at which the light from the bright spot is beamed, measured in the same way as $\phi$.}
\label{tab:mcmcresults} 
\begin{tabular}{lr@{\,$\pm$\,}lr@{\,$\pm$\,}lr@{\,$\pm$\,}l} 
\hline
           & \multicolumn{2}{c}{$u'$-band}  & \multicolumn{2}{c}{$g'$-band}  & \multicolumn{2}{c}{$r'$-band}\\
\hline
$q$                 & 0.457 & 0.005                 & 0.481 & 0.003                 & 0.497 & 0.005\\
$i$                 & 84.08 & 0.20                  & 83.64 & 0.10                  & 83.31 & 0.16 \\
$R_1/a$               & 0.0058 & 0.0004               & 0.0045 & 0.0002               & 0.0051 & 0.0003 \\ 
$R_{disc}/a$          & 0.333 & 0.004                 & 0.285 & 0.006                 & 0.281 & 0.004 \\
$R_{spot}/a$          & 0.263 & 0.001                 & 0.261 & 0.001                 & 0.257 & 0.0001\\
$\beta$      & 0.51  & 0.06                  & 0.19  & 0.02                  & 0.25  & 0.03\\
$\gamma$       & 0.727 & 0.022                 & 0.532 & 0.006                 & 0.505 & 0.009 \\
$\delta$      & 0.041 & 0.069                 &-0.262 & 0.065                 &-0.100 & 0.045 \\
$l$     & 0.0102 & 0.0012               & 0.0097 & 0.0001               & 0.0096 & 0.0001\\
$f_c$      & 0.711 & 0.008                 & 0.723 & 0.004                 & 0.723 & 0.006\\
$\phi$      & 105.9 & 1.6                   & 126.6 & 0.5                   & 125.1 & 0.6 \\
$\psi$        & 59.1 & 1.7                    & 32.7 & 0.8                    & 39.6 & 0.8 \\
\hline
\end{tabular}
\end{table}

\end{document}